\documentclass [prb,preprint,superscriptaddress,floatfix,amsmath,amssymb] {revtex4-1}
\usepackage{amsmath,amssymb}
\usepackage{graphicx}
\usepackage{txfonts}
\usepackage{color}
\usepackage{hyperref}
\usepackage{adjustbox}
\usepackage{bbm}
\usepackage{framed}
\usepackage{wrapfig}
\usepackage{multirow}
\usepackage[letterpaper]{geometry}
\global\long\def\av#1{\left\langle #1 \right\rangle }

\renewcommand{\vec}[1]{\boldsymbol{#1}}
\newcommand{\beq}{\begin{equation}}
\newcommand{\eeq}{\end{equation}}
\newcommand{\bea}{\begin{eqnarray}}
\newcommand{\eea}{\end{eqnarray}}

\begin{document}

\title{
Valley magnetism, nematicity, and density wave orders in twisted bilayer graphene}

\author{Dmitry V. Chichinadze}
\affiliation{School of Physics and Astronomy,
University of Minnesota, Minneapolis, MN 55455, USA}
\author{Laura Classen}
\affiliation{School of Physics and Astronomy,
University of Minnesota, Minneapolis, MN 55455, USA}
\affiliation{Condensed Matter Physics \& Materials Science Division, Brookhaven National Laboratory, Upton, New York 11973, USA}
\author{Andrey V. Chubukov}
\affiliation{School of Physics and Astronomy,
University of Minnesota, Minneapolis, MN 55455, USA}
\begin{abstract}
We analyze density-wave and Pomeranchuk orders in twisted bilayer graphene.
 This compliments our earlier analysis of the pairing instabilities.  We assume that near half-filling of either conduction or valence band, the Fermi level is close to Van Hove points, where the density of states diverges, and study potential instabilities in the particle-hole channel within a patch model with two valley degrees of freedom. The hexagonal symmetry of twisted bilayer graphene allows for  either six or twelve Van Hove points. We consider both cases and find the same two leading candidates for particle-hole order. One is an SU(2)-breaking spin state with ferromagnetism within a valley.
 A subleading inter-valley hopping induces antiferromagnetism between the valleys.
The same state has also been obtained in strong coupling approaches, indicating that this order is robust.
The other is a mixed state with $120^\circ$ complex spin order and orthogonal complex charge order.
 In addition, we find a weaker, but still attractive interaction in nematic channels, and discuss the type of a nematic order.
     \end{abstract}

\maketitle

\section{Introduction}
\label{sec:Intro}

The discovery of superconductivity \cite{Cao2018unconventional} and correlated insulating states \cite{Cao2018correlated} in magic-angle twisted bilayer graphene (TBG) has generated an enormous interest in the physics of
this \cite{Yankowitz2019,Lu2019,stm2019,Cao2020nematicity,Jiang2019,Choi_2019,2019Natur572101X,tschirhart2020imaging,Serlin900,Sharpe605,Saito_2020,Polshyn2019,liu2020tuning} and related systems \cite{2019arXiv190308596C,2019arXiv190308130L,2019arXiv190306952S,trilayer1,trilayer2,chen2020,polshyn2020nonvolatile}.  A lot of effort, both experimental and theoretical, is put forward to understand the underlying mechanism of superconductivity and strong correlations \cite{Young2020}.

An essential question in this context is the ratio of the interaction and the fermionic bandwidth, and the associated appropriate theoretical framework.
Experimental data indicate that the effective electron-electron interaction in magic-angle TBG  is comparable to the bandwidth \cite{stm2019}, similar to the case of cuprate superconductors.
By this reason, the physics of TBG has been studied within both strong-coupling \cite{Balents2018,Po2018PRX,Kang2018strong,Thomson2018PRB,Bultinck2019ground,Zhang2019,Khalaf2020charged,Ledwith2020,Chatterjee2020,
Bultinck2020,Repellin2020,Repellin2020PRL,Zhang2020quantum,Xie2020,Zhu2020curious,PhysRevLett.122.246402,Bultinck2020,PhysRevB.98.121406}
and itinerant
\cite{Kivelson2018,Koshino2018PRX,Isobe2018PRX,Lin2018,Gonzalez2019,Lin2019,Chichinadze2019nematic,PhysRevLett.121.217001,Kozii2020superconductivity,PhysRevB.98.241407,PhysRevB.99.195120,You2019npj,Wolf2019,Sboychakov2018}  approaches.
Strong coupling approaches assume that correlated phases are some versions of Mott insulators and can be understood by taking interactions to be much larger than the bandwidth.
Itinerant approaches assume that low-energy physics can be analyzed by focusing on a subset of states near the Fermi surface, and that both superconductivity and correlated phases can be understood as instabilities of a Fermi liquid in particle-particle and particle-hole channels.

One robust feature of TBG, detected by scanning tunnelling spectroscopy and Hall density measurements \cite{e_andrei,stm2019,wu2020chern}, is the existence of sharp peaks in the density of states. These peaks are often interpreted as originating from Van Hove points \cite{PhysRev.89.1189} -- the saddle points in the electron dispersion.
   Tight-binding models for the electron dispersion of TBG (Refs. \cite{Yuan2018,Kang2018PRX,Isobe2018PRX}) do possess Van Hove points, and these points are located near the Fermi level at half-filling of both hole and electron 
      bands
   ($n= \pm 2$) and, possibly, at
   $n=\pm 3$\cite{wu2020chern} (in the classification when the full bandwidth is between $n=4$ and $n =-4$).
    Near $n = \pm 2$, the number of Van Hove points is either six or twelve, depending on the hopping parameters.  The presence of Van Hove points generally increases the strength of correlation effects.
 This has been used as an argument that the observed superconductivity and correlated behavior near $n = \pm 2$ may be due to Van Hove physics.

In our previous study~\cite{Chichinadze2019nematic}, we analyzed pairing instabilities within the effective models for six and twelve Van Hove points. For the model with six Van Hove points, we reproduced earlier results~\cite{Isobe2018PRX,Lin2018,PhysRevB.98.085436,PhysRevLett.121.217001,Lin2019,PhysRevB.98.241407,PhysRevB.99.195120,You2019npj}
 that the ground state has a chiral $d \pm id$ superconducting order, which breaks time-reversal symmetry, but leaves the lattice rotation symmetry intact. For twelve Van Hove points, we found two attractive channels, $g$ and $i$-waves, with almost equal coupling constants, and showed that in the coexistence state the threefold lattice rotation symmetry is broken, i.e., the superconducting state is also a nematic. We argued that this is consistent with the experimental data near $n =-2$ (Ref.\cite{Cao2020nematicity}).

In this paper, we analyze potential instabilities in the particle-hole channel and the corresponding free energies.
We determine the effective couplings in various spin-density wave (SDW), charge-density wave (CDW), and
 spin and charge Pomeranchuk channels (i.e., particle-hole channels with zero momentum transfer),  find which channels are attractive and in which one the attractive coupling is the strongest.

We investigate the  leading instabilities in the particle-hole channel using the real-space interaction Hamiltonian suggested by Kang and Vafek~\cite{Kang2018strong}.
This Hamiltonian has two terms.  One is a cluster Hubbard term, which contains density-density interactions between sites of a given hexagon in the moiré lattice. The second term is a bilinear combination of hoppings between different sites of a hexagon.
It includes terms that are often called pair hopping and exchange interactions, again between all sites of a hexagon.
The relative strength of the two terms is parametrized by a dimensionless $\alpha_T$ (see below), which was argued to be of order one~\cite{Kang2018strong}.  Here, we use $\alpha_T$ as an input parameter.
We convert the interaction into momentum space, project onto the vicinity of the Van Hove points and analyze the dressed couplings in different channels
  for $0 \leq \alpha_T \leq 1$.

Particle-hole instabilities in the vicinity of Van Hove points in TBG have been studied previously for the six-patch model and $\alpha_T =0$ (Refs. \cite{Isobe2018PRX,Gonzalez2019,Lin2019,Lu2020chiral}).
It was argued that the leading instability is degenerate between SDW and CDW and occurs at all three degenerate symmetry-related momenta that connect
the six Van Hove points. We found the same instability in the six-patch model in some range of finite $\alpha_T$. We go beyond earlier studies and derive and analyze the corresponding free energies to determine the actual composition of the order parameter. We argue that the ground state is a mixed SDW/CDW state with three-component, complex SDW and CDW orders, ${\bf m_i} e^{i\phi_i}$ and $\Delta_i e^{i \psi_i}$, $i=1,2,3$.  The spin components ${\bf m}_i$  form a $120^{\circ}$ configuration and  the phase difference between charge and spin components is $\psi_i -\phi_i = \pm \pi/2$ for all $i$.   This state breaks translational and time-reversal symmetry.

For larger $\alpha_T$ in the six-patch model and for all $\alpha_T$ in the twelve patch model, we find the leading instability in the  s-wave spin Pomeranchuk channel. The corresponding order is
 O(3)
ferromagnetism within a given valley.
The relative orientation of the magnetic moments in the two valleys depends on
 the interplay between weaker subleading terms.
  We find that inter-valley hopping terms favor
 antiferromagnetism between the two valleys.
We label this state as FM/AFM. It is also called a valley antiferromagnet.
The same FM/AFM order has been obtained in the strong coupling approach. Kang and Vafek found this order near half-filling \cite{Kang2018strong}.
 Other groups found FM/AFM order also at different fillings \cite{PhysRevB.100.205131,PhysRevResearch.2.013370}.
 The emergence of the same FM/AFM state in both  itinerant and strong coupling approaches is an indication that this order is rather robust and likely not very sensitive to the closeness to the Van Hove filling (for a similar discussion for bilayer graphene see Ref. \cite{Vafek2010weakstrong}). The SDW/CDW state has not been detected at strong coupling.

We also analyze interactions in non-s-wave Pomeranchuk channels. We argue that the interaction in the $d$- or $g$-wave
 charge and spin channels (depending on the model) is attractive, even when $\alpha_T =0$. We argue that this is a consequence of the fact that the cluster Hubbard interaction contains terms with the products of electronic densities at different sites of a hexagon.
For only an on-site Hubbard interaction, the couplings
 in $d$- or $g$-wave Pomeranchuk channels would either be repulsive or vanish \cite{Xing2017PRB}.
We argue on general grounds that these instabilities give rise to nematicity, i.e., a
 non s-wave
Pomeranchuk order breaks lattice rotational symmetry.

 Within our model, bare interactions in the non-s-wave Pomeranchuk channels are  subleading to that in the FM/AFM channel. However, the strength of the interaction in different channels varies as one progressively integrates our high energy fermions, it is possible that an attraction in a nematic channel may exceed those in other particle-hole channels.  With this in mind, and also motivated by the experiments which show evidence for strong nematic fluctuations and, possibly, a nematic order in the normal state for some dopings~\cite{stm2019,Cao2020nematicity,Jiang2019}, we analyze what kind of nematic order can emerge in both six- and twelve-patch models.

The structure of the paper is the following. In the next section we briefly discuss the evolution of the Fermi surface in TBG away from charge neutrality and introduce six- and twelve-patch models. The corresponding Hamiltonians include all possible scattering processes between low-energy fermions. We express the coupling constants via the parameters of the underlying lattice model, which contains extended density-density and exchange interactions within the honeycombs of the moir\'e  superlattice \cite{Kang2018strong}. The relative strength of the exchange interactions is specified by the dimensionless $\alpha_T$.
In Sec.~\ref{sec:RPA6patch} we analyze spin and charge orders in the six-patch model. We introduce trial particle-hole vertices with zero momentum transfer and with momentum transfers equal to the distance between Van Hove points. We obtain the set of coupled equations for the full vertices within the ladder approximation, and extract the couplings in each particle-hole channel.  We identify the subset of channels for which the couplings are attractive and show that the ones in the SDW/CDW and FM/AFM channels are the most attractive, followed by those in the $d$-wave Pomeranchuk channels.
      In each case, the leading eigenvalue is degenerate.
 In Sec.~\ref{sec:Landau6patch} we derive the Landau functional for the SDW/CDW and the FM/AFM order parameters and in each case determine the actual order-parameter configuration.
   In Sec.~\ref{sec:dPoms} we discuss the Landau functional for $d$-wave spin and charge Pomeranchuk
  order parameters and argue that the corresponding orders break lattice rotational symmetry.
In Sec.~\ref{sec:12patch} we perform the same analysis as in Secs. \ref{sec:RPA6patch}- ~\ref{sec:dPoms} for the twelve-patch model.
   We present our conclusions in Sec.\ref{sec:Concl}.

\section{The patch model}
\label{sec:TBM}

As we said in the introduction, the measured  density of states of TBG shows peaks at around half-filling of conduction and valence bands. The most natural explanation for the peaks is the presence of the Van Hove saddle points in the electronic dispersion.  At the quasiparticle energy where the Van Hove points lie at the Fermi level, the dispersion undergoes a topological change (Lifshitz transition), and the density of states shows a spike. Van Hove saddle points generally appear in two dimensional materials as a consequence of the periodicity of the energy dispersion \cite{PhysRev.89.1189}. Because of the rotational symmetry of TBG\cite{Venderbos2018}, the number of Van Hove points is a multiple of six. Earlier analysis of  tight-binding models have found that there can be either six or twelve Van Hove points \cite{Yuan2018,Koshino2018PRX,Kang2018PRX,Yuan2019magic,Gonzalez2019}

\begin{figure}[h!]
\includegraphics[width=0.95\linewidth]{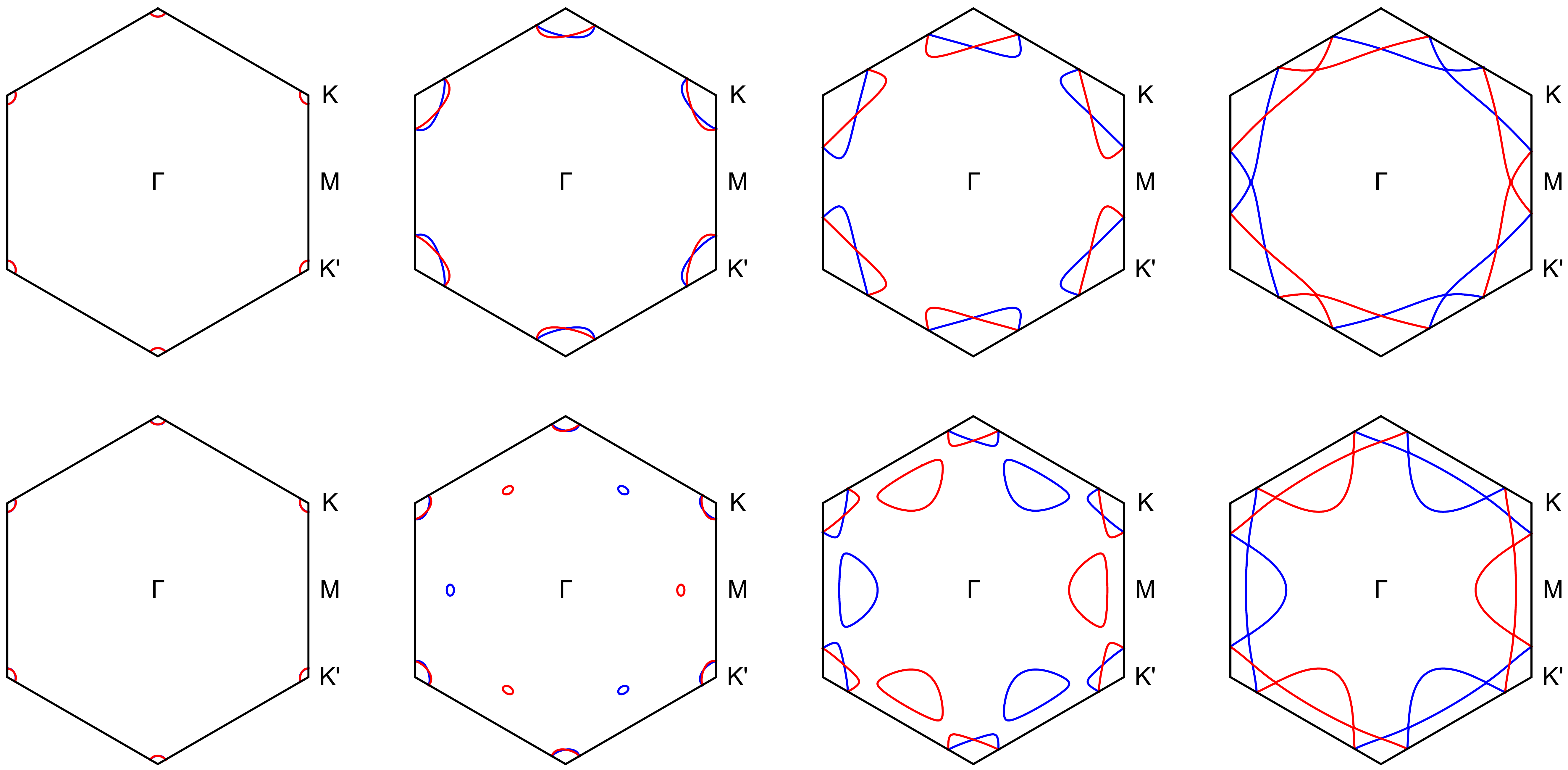}
\caption{The evolution of Fermi surface upon doping towards 6 Van Hove points (upper panel) and 12 Van Hove points (lower panel). }
\label{fig:fsevol}
\end{figure}

In the vicinity of Van Hove filling, i.e. when the Fermi level lies near the Van Hove energy, the density of states is enhanced and amplifies the  effects of the interactions between fermions in patches around the Van Hove points. The interactions may give rise to an instability of the Fermi liquid already for moderate couplings.
This situation can be described in terms of patch models which consider the most general Hamiltonian for fermions around the Van Hove points.
 We emphasize that patch models can be rationalized without reference to a particular tight-binding dispersion as Van Hove points should necessarily be present in any model that contains Dirac points at charge neutrality and a Fermi surface centered at $\Gamma$ at large hole or electron doping.
 For concreteness, we give an example in the next section based on the tight-binding model of Refs.~\onlinecite{Yuan2018,Koshino2018PRX}. The corresponding evolution of the Fermi surface is shown in Fig.~\ref{fig:fsevol}.

\subsection{Effective patch models near Van Hove points in TBG from tight-binding Hamiltonian}

The tight-binding Hamiltonian of Refs.~\onlinecite{Yuan2018,Koshino2018PRX} for electrons on the moir\'e superlattice is given by
\begin{align}\label{2H}
H_{TB} &= - \mu\sum_i\sum_{o=\pm}  c_{io}^{\dagger} c_{io} +  t_1 \sum_{\av{ij}} \sum_o\left[c_{io}^{\dagger} c_{jo} + h.c. \right] +  t_2\sum_{\av{ij}_5} \sum_o  \left[c_{io}^{\dagger} c_{jo} + h.c. \right] \\
- i t_3 \sum_{\av{ij}_5} \left[ c_{i+}^\dagger c_{j,+} -c_{i-}^\dagger c_{j-} + h.c. \right].
\end{align}
The sums go over the sites,  which represent the AB or BA regions of the honeycomb moir\'e superlattice in TBG, $\mu$ denotes the chemical potential, $t_1, t_2$ are real hopping amplitudes between nearest- and fifth-nearest-neighbors, and $\av{ij}_5$ denotes fifth-nearest neighbors with hopping amplitude $t_3$. A fifth-nearest neighbor is equivalent to a second-nearest neighbor within the same sublattice. The electron annihilation operators $c_{io} $ possess
an index $o$,  inherited from the valleys of the original graphene sheets.
This index is sometimes  called as orbital index and sometimes valley index.  For definiteness, we will use the "valley" notation.
We label two values of $o$ as $+$ and $-$.

 The Hamiltonian is spin SU(2) symmetric, and spin indexes are suppressed for simplicity. It also possesses time-reversal and valley U(1) symmetry, which can be traced back to the suppression of inter-valley coupling in small-angle TBG \cite{Koshino2018PRX}.  The space symmetry of the TBG lattice is described by the group $D_3$ \cite{Koshino2018PRX} (see also \cite{Venderbos2018}).

\begin{figure}[t]
\includegraphics[width=0.22\linewidth]{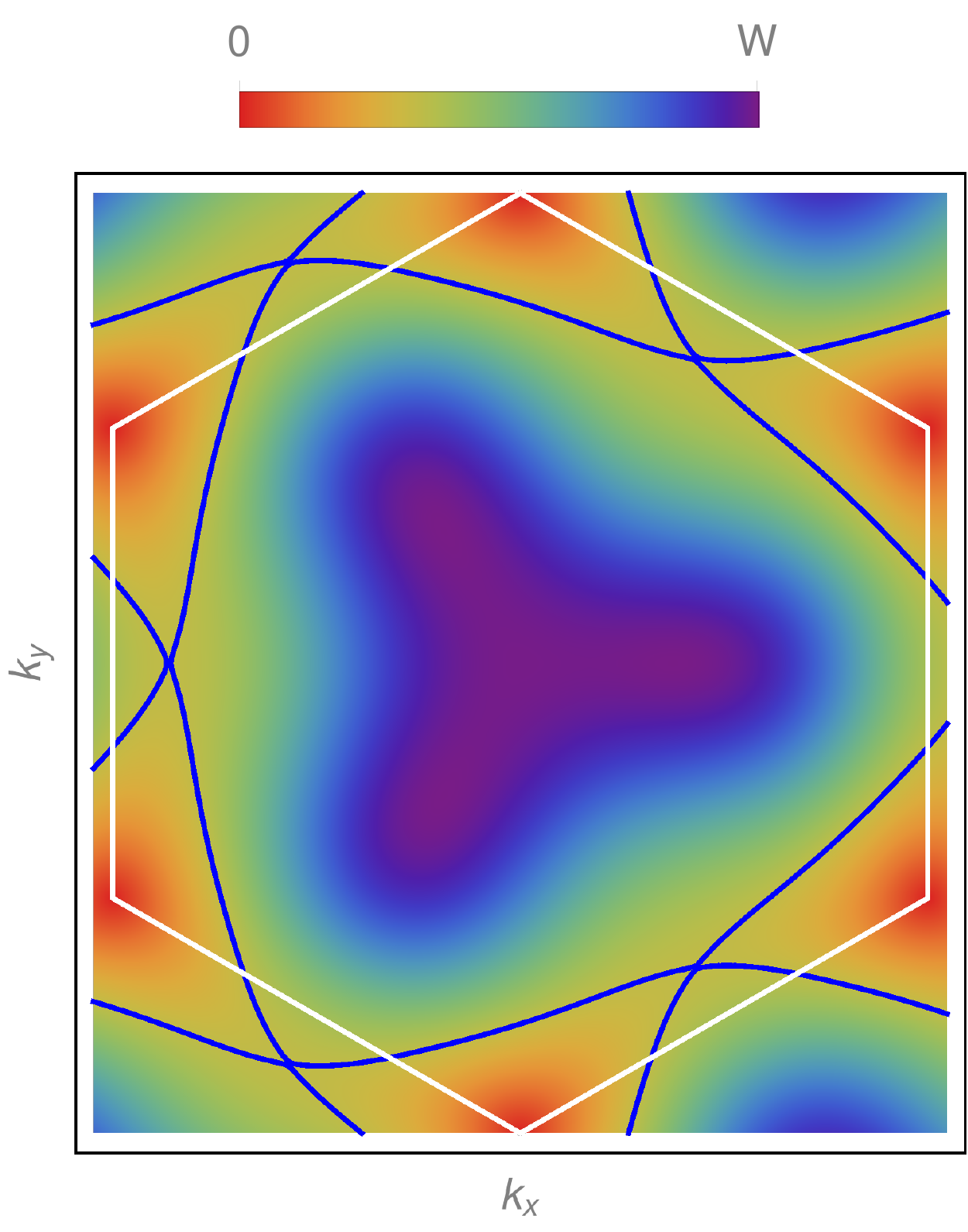} \quad
\includegraphics[width=0.22\linewidth]{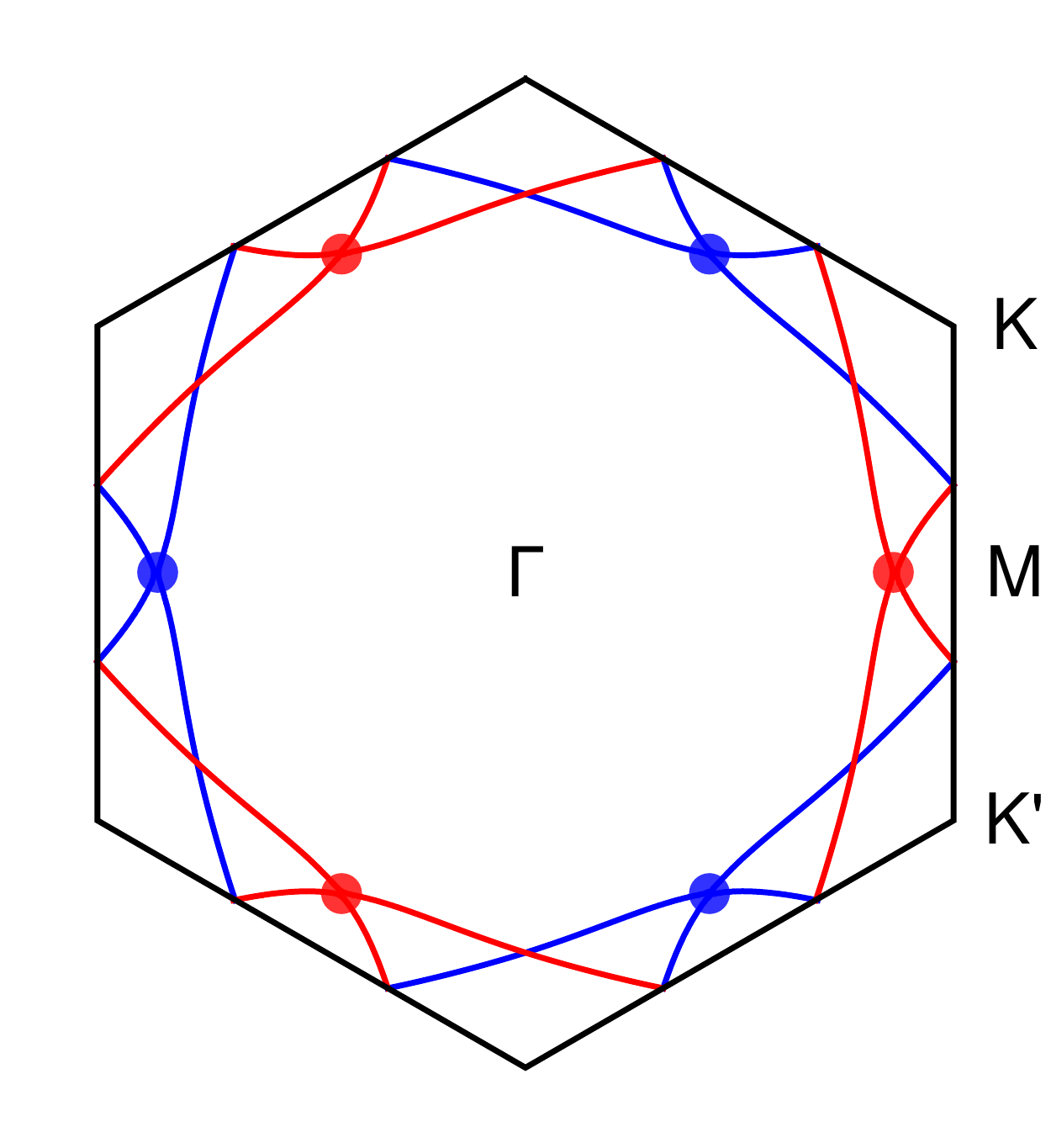} \qquad
\includegraphics[width=0.22\linewidth]{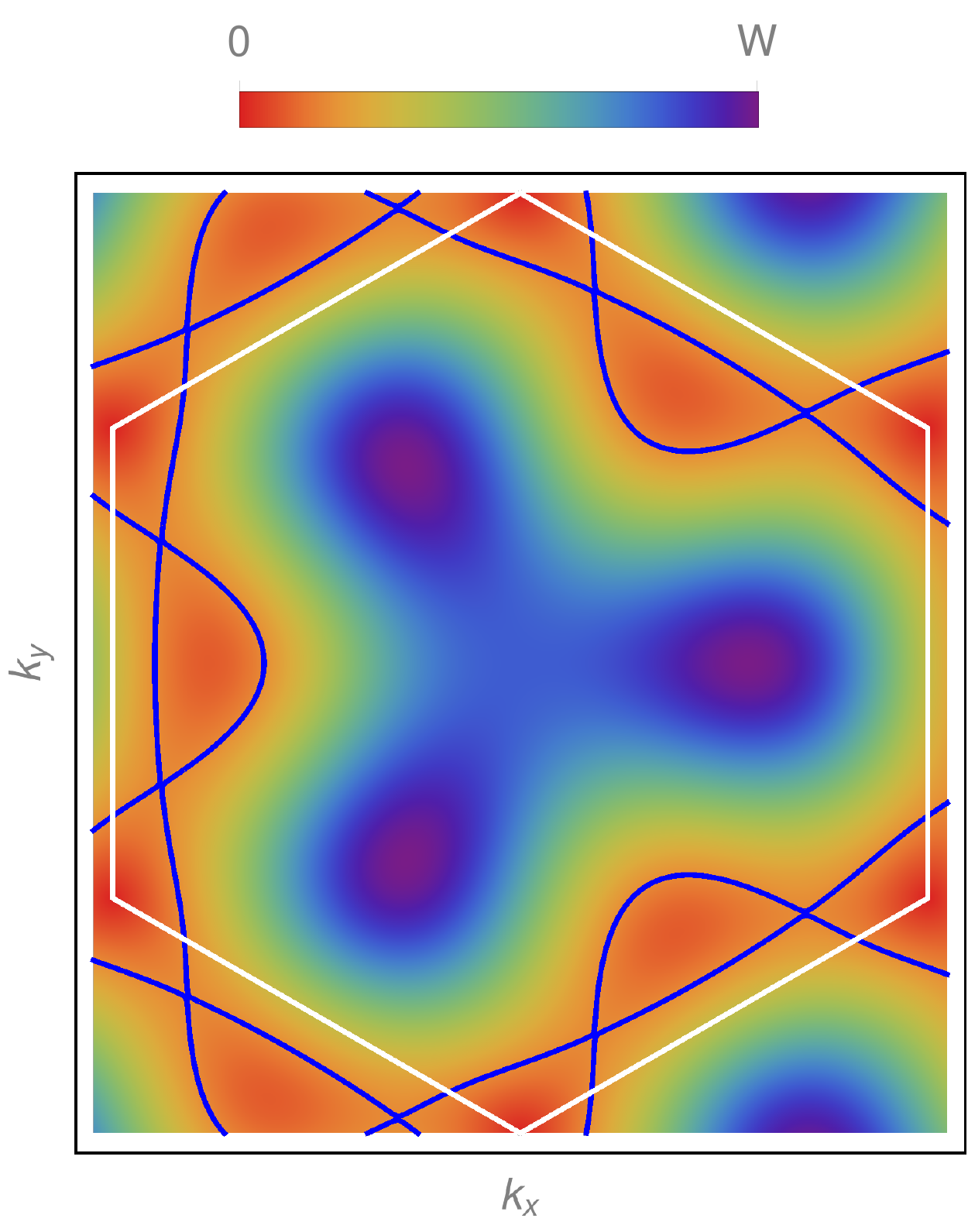} \quad
\includegraphics[width=0.22\linewidth]{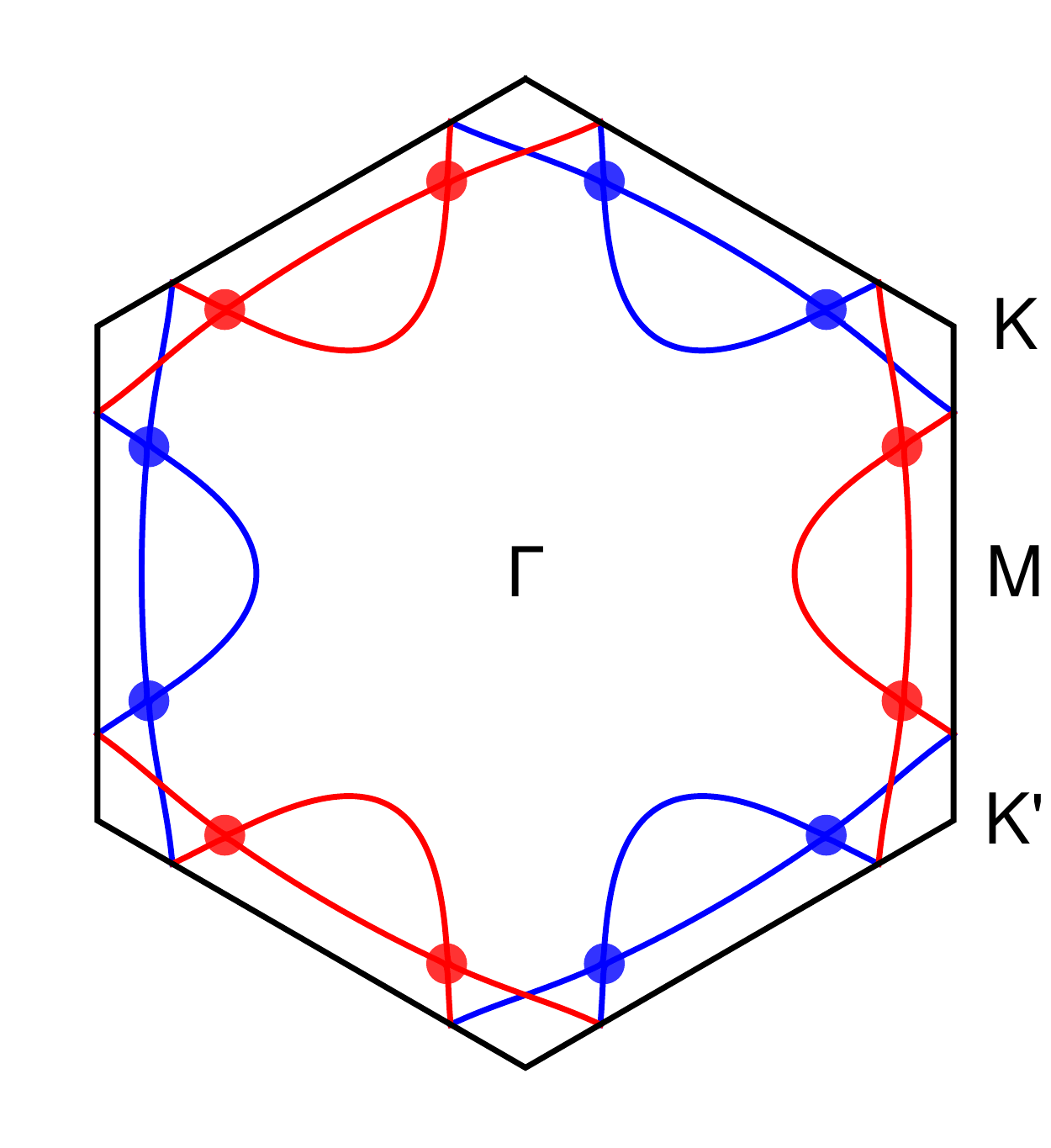}
\caption{Examples of the Fermi surface with six (left) and twelve (right) Van Hove points. The color encodes the energy dispersion of one of the two conduction bands from charge neutrality to the bandwidth $W$.
The corresponding Fermi surface is shown in blue in the color plot and the sketches to their right. The total Fermi surface also contains the contribution from the second band shown in red. Van Hove points are marked by blue and red disks.
The two conduction (or the two valence) bands are valley polarized, i.e. they belong to either the $+$ (red) or $-$ (blue) valley.
}
\label{fig:dispersion}
\end{figure}

In momentum space, Hamiltonian \eqref{2H} yields two spin-degenerate valence and conduction bands.
\begin{align}
E^v_\pm &= -|T_{sd1}|+T_d \pm T_{sd2} -\mu\\
E^c_\pm &= +|T_{sd1}|+T_d \pm T_{sd2} -\mu,
\label{band_spec}
\end{align}
where
\begin{align}
T_d &=-\mu+2 t_2 \left(\cos{ \frac{3}{2} \left(-k_x + \sqrt{3} k_y \right)} + \cos{\frac{3}{2} \left(-k_x - \sqrt{3} k_y \right)} +\cos{ 3 k_x} \right), \\
T_{sd1} &= t_1 \left( \exp(i k_x) + 2 \exp(-i \frac{k_x}{2}) \cos({\frac{\sqrt{3} k_y}{2}}) \right), \\
T_{sd2} &= 2 t_3 \left(\sin{\frac{3}{2} \left(-k_x + \sqrt{3} k_y \right)} + \sin{\frac{3}{2} \left(-k_x - \sqrt{3} k_y \right)} +\sin{3 k_x} \right).
\end{align}
 The bands are valley polarized, i.e. there is no hybridization between $+$ and $-$ valleys.
  Even in this case, the transformation to the bands is still non-trivial because of the sublattice degrees of freedom.
   The bands possess Van Hove points, whose number can be six or twelve,  depending on the hopping amplitudes.
   Each of the two bands
    contributes half of the Van Hove points.
    We show
 the dispersion for both cases in Fig.~\ref{fig:dispersion}.
  In the case of six Van Hove points, they lie along the $\Gamma$-M line and symmetry-related directions, at some distance  from the zone boundary. When twelve Van Hove points are present, they do not lie along any symmetry direction in the Brillouin zone.

   Upon electron or hole doping, the energies, at which Van Hove points are located, move closer to Fermi energy and cross it at particular doping levels. We consider system behavior near these particular dopings,
  focus on the low-energy states,
   and introduce effective patch models with momenta in a finite range near either six or twelve Van Hove points.  To this end, we expand the energies around the Van Hove points and approximate the hopping Hamiltonian by
\begin{align}
\label{eq:Hpatch0}
H=\sum_{i=1}^{N_p} \sum_{\sigma=\uparrow,\downarrow} \left[ \epsilon_{i}(\vec k) f_{i\sigma}^\dagger(\vec k) f_{i\sigma}(\vec k) + \epsilon_{i'}(\vec k) f_{i'\sigma}^\dagger(\vec k) f_{i'\sigma}(\vec k)\right],
\end{align}
where $f_{i\sigma}(\vec k)$ ($f_{i'\sigma}(\vec k)$) describes an electron from a given valley in the vicinity of patch $i$ with momentum $\vec k$ and spin $\sigma$. The patch index $i$ runs from $1$ to $N_p$. For the case of six patches $N_p=3$ (three patches for fermions from each of the two valleys $o=\pm$, which we also label by the addition of a prime or no prime in the following), for twelve patches $N_p=6$, see Fig. \ref{mod_sk}.  The dispersions $\epsilon_{i(i')}(\vec k)$ have hyperbolic forms. Within one band, $\epsilon_{i}(\vec k)$ and $\epsilon_{j}(\vec k)$ are related by $D_3$ symmetry, while $\epsilon_{i}(\vec k)$ and $\epsilon_{i'}(\vec k)$ are related by time-reversal symmetry (see Fig.~\ref{fig:dispersion}).

\subsection{Couplings in the 6-patch model}

\begin{figure}[t]
\includegraphics[width=0.8\linewidth]{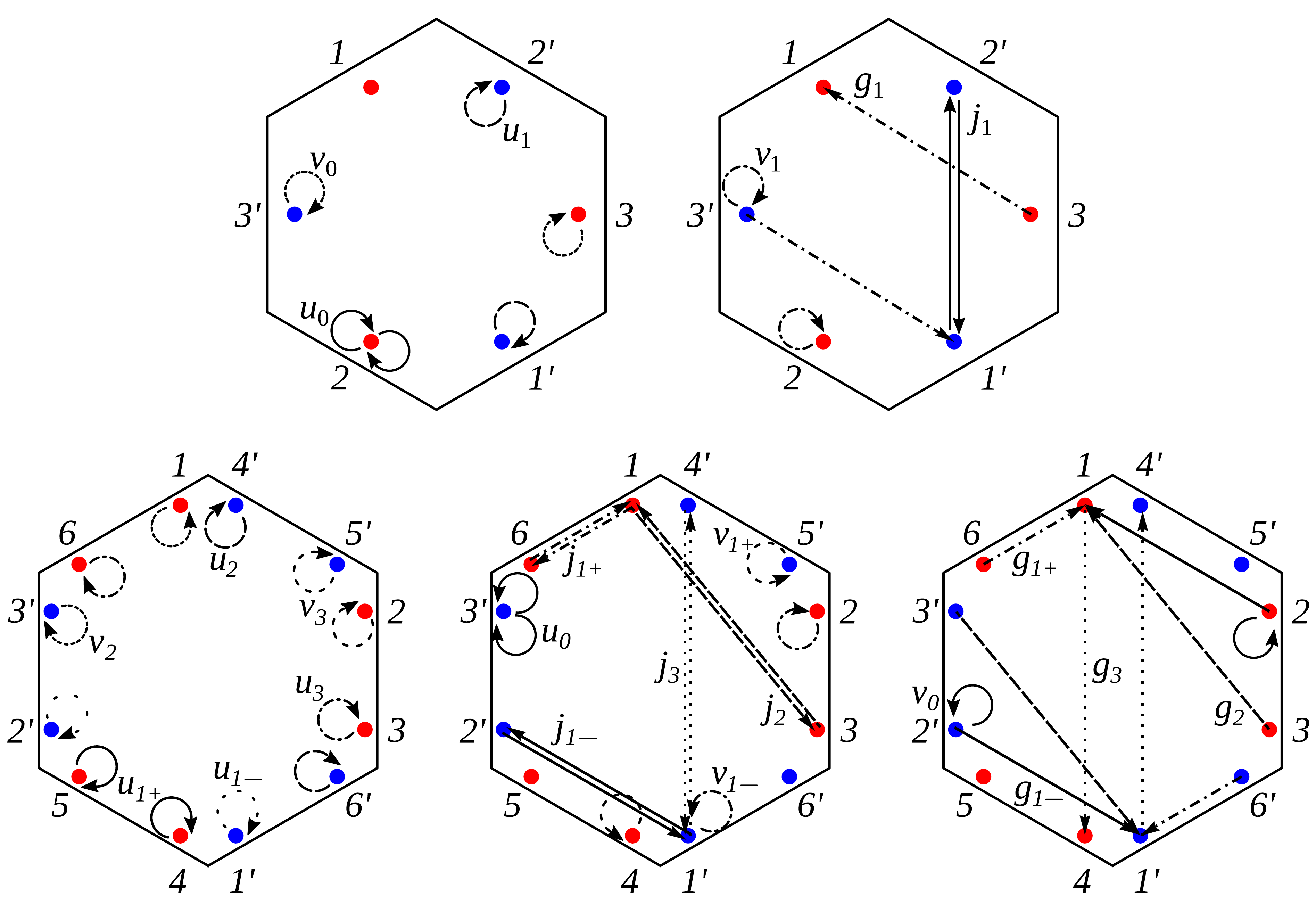}
\caption{ Sketch of the interactions in six-patch (upper line) and twelve-patch model (lower line). Blue (red) dots mark the Van Hove points made of valley $o=-(+)$. We label the patches $i=1\ldots N_p$ with $N_p=3$ for the six-patch and $N_p=6$ for the twelve patch model. We distinguish if a patch is made from valley $+$ or $-$ via adding a prime to the patch number or not ($i$ vs. $i'$).}
\label{mod_sk}
\end{figure}

We next consider all symmetry-allowed couplings between fermions within the six patches, with the restriction that we exclude valley mixing terms.  Valley mixing terms are interaction processes of the form $f_{i\sigma}^\dagger f_{j'\sigma} f_{i'\sigma'}^\dagger f_{j\sigma'}$ that involve different valleys. These terms are present~\cite{Isobe2018PRX}, but were found to be very small numerically   ~\cite{Kang2018strong,Koshino2018PRX}.
In general, there are six  different intra- and inter-patch density-density and exchange interactions \cite{Isobe2018PRX,Chichinadze2019nematic}
\begin{align}
H^{Int}_{6p}&= \sum_{i=1}^3 \left[  u_0 \left( f_{i}^\dagger f_{i} f_{i}^\dagger f_{i} + f_{i'}^\dagger f_{i'} f_{i'}^\dagger f_{i'} \right) + v_0
f_{i}^\dagger f_{i} f_{i'}^\dagger f_{i'}
+ u_1 \left( f_{i}^\dagger f_{i} f_{i+1}^\dagger f_{i+1} +f_{i'}^\dagger f_{i'} f_{(i+1)'}^\dagger f_{(i+1)'} \right)  \right.
\notag \\
&\left.  + v_1 \left( f_{i}^\dagger f_{i} f_{(i+1)'}^\dagger f_{(i+1)'} + f_{i'}^\dagger f_{i'} f_{i+1}^\dagger f_{i+1} \right)
 + j_1 \left( f_{i}^\dagger f_{i+1} f_{i+1}^\dagger f_{i} + f_{i'}^\dagger f_{(i+1)'} f_{(i+1)'}^\dagger f_{i'} \right) +  g_1 \left ( f_{i}^\dagger f_{i+1} f_{i'}^\dagger f_{(i+1)'} + \text{h.c.}\right)\right].
\label{eq:iap6}
\end{align}
 The spin structure of  every  term is $\sum_{\sigma,\sigma'}f_{\sigma}^\dagger f_\sigma f_{\sigma'}^\dagger f_{\sigma'}$. The six scattering processes $u_0,v_0,u_1,v_1, j_1$ and $g_1$  are sketched in Fig. \ref{mod_sk}.
Umklapp processes are forbidden because Van Hove singularities do not appear at momenta connected by a reciprocal lattice vector.
 If we treat Eq.~(\ref{eq:iap6}) as the effective low-energy model, which incorporates the renormalizations of the interactions by fermions outside the patches, then the interactions depend on the transferred momenta,
the total incoming momenta, and the exchanged momenta (the transferred momenta in the antisymmetrized vertex, with outgoing fermions interchanged), and we have to treat all six interactions as different.
 In this paper we use the bare values of the interactions. In this case, the couplings are functions of momentum transfer only, and we have $u_0 = u_1=v_0 = v_1= u$ and $g_1 = j_1 =g.$

\subsection{Couplings in the 12-patch model}

In the case of the twelve-patch model, there are more symmetry-allowed interaction processes.
 Without valley mixing we obtain \cite{Chichinadze2019nematic}
\begin{align}
H^{int}_{12p} &= \sum_{i=1}^6 \left[ u_0 \left( f_{i}^\dagger f_{i} f_{i}^\dagger f_{i} + f_{i'}^\dagger f_{i'} f_{i'}^\dagger f_{i'} \right) +   v_0
f_{i}^\dagger f_{i} f_{i'}^\dagger f_{i'}
+ u_2 \left( f_{i}^\dagger f_{i} f_{i+2}^\dagger f_{i+2} +  f_{i'}^\dagger f_{i'} f_{(i+2)'}^\dagger f_{(i+2)'} \right) \right. \notag \\
& \left.  +  v_2 \left( f_{i}^\dagger f_{i} f_{(i+2)'}^\dagger f_{(i+2)'} +  f_{i'}^\dagger f_{i'} f_{i+2}^\dagger f_{i+2} \right) + u_3 \left( f_{i}^\dagger f_{i} f_{i+3}^\dagger f_{i+3} + f_{i'}^\dagger f_{i'} f_{(i+3)'}^\dagger f_{(i+3)'}  \right) +  v_3 \left( f_{i}^\dagger f_{i} f_{(i+3)'}^\dagger f_{(i+3)'} + f_{i'}^\dagger f_{i'} f_{i+3}^\dagger f_{i+3}  \right) \right. \notag \\
& \left. + j_2 \left( f_{i}^\dagger f_{i+2} f_{i+2}^\dagger f_{i} + f_{i'}^\dagger f_{(i+2)'} f_{(i+2)'}^\dagger f_{i'} \right) + g_2 \left( f_{i}^\dagger f_{i+2} f_{i'}^\dagger f_{(i+2)'} + \text{h.c.} \right)+ j_3 \left( f_{i}^\dagger f_{i+3} f_{i+3}^\dagger f_{i} + f_{i'}^\dagger f_{(i+3)'} f_{(i+3)'}^\dagger f_{i'} \right)  \right. \notag \\
&\left. + g_3 \left( f_{i}^\dagger f_{i+3} f_{i'}^\dagger f_{(i+3)'} + \text{h.c.} \right)  + u_{1+} \left( f_{i}^\dagger f_{i} f_{i+(-1)^i}^\dagger f_{i+(-1)^i} + f_{i'}^\dagger f_{i'} f_{i'+(-1)^{i'}}^\dagger f_{i'+(-1)^{i'}} \right) \right. \notag \\
&\left.+ u_{1-} \left( f_{i}^\dagger f_{i} f_{i-(-1)^i}^\dagger f_{i-(-1)^i} + f_{i'}^\dagger f_{i'} f_{i'-(-1)^{i'}}^\dagger f_{i'-(-1)^{i'}} \right)  \right. \notag \\
& + v_{1+} \left( f_{i}^\dagger f_{i} f_{i'+(-1)^{i'}}^\dagger f_{i'+(-1)^{i'}} + f_{i'}^\dagger f_{i'} f_{i+(-1)^{i}}^\dagger f_{i+(-1)^{i}} \right) + v_{1-} \left( f_{i}^\dagger f_{i} f_{i'-(-1)^{i'}}^\dagger f_{i'-(-1)^{i'}} + f_{i'}^\dagger f_{i'} f_{i-(-1)^i}^\dagger f_{i-(-1)^i} \right)  \notag \\
&\left.
+ j_{1+} \left( f_{i}^\dagger f_{i+(-1)^i} f_{i+(-1)^i}^\dagger f_{i} + f_{i'}^\dagger f_{i'+(-1)^{i'}} f_{i'+(-1)^{i'}}^\dagger f_{i'}  \right)  + g_{1+}  \left( f_{i}^\dagger f_{i+(-1)^i} f_{i'}^\dagger f_{i'+(-1)^{i'}} + \text{h.c.} \right) \right. \notag \\
&\left. + j_{1-} \left( f_{i}^\dagger f_{i-(-1)^i} f_{i-(-1)^i}^\dagger f_{i} + f_{i'}^\dagger f_{i'-(-1)^{i'}} f_{i'-(-1)^{i'}}^\dagger f_{i'} \right) + g_{1-} \left( f_{i}^\dagger f_{i-(-1)^i} f_{i'}^\dagger f_{i'-(-1)^{i'}}+ \text{h.c.} \right) \right]\,
\label{eq:iap12}
\end{align}
We again suppressed the spin index for simplicity, each term is of the form $\sum_{\sigma,\sigma'}f_{\sigma}^\dagger f_\sigma f_{\sigma'}^\dagger f_{\sigma'}$.
We sketch the couplings in Fig.~\ref{mod_sk}.   In general, there are 18 different couplings. We assume, as before that the interactions are the bare ones, and  depend only on the momentum transfer. In this case, there are five independent couplings
\begin{equation}
\begin{gathered}
u_0 = u_{1-} = u_{1+} = u_2 = u_3 = v_0 = v_{1-} = v_{1+} = v_2 = v_3 = u; \\
j_{1-} = g_{1-}; \; \; j_{1+} = g_{1+}; \; \;  j_2 = g_2; \; \; j_3 = g_3.
\end{gathered}
\end{equation}

\subsection{Bare values of the couplings -- comparison with the non-local microscopic model}

The bare values for the couplings in the patch models can be obtained by choosing a particular microscopic model
and projecting microscopic interactions onto the patches. We use the model of Ref. \cite{Kang2018strong}, which includes the cluster Hubbard density-density interaction and the bi-products of hoppings between fermions within a given hexagon:
\begin{equation}
H_{int} =
V_0 \sum_{\bold{R}} \left( \sum_{o=\pm} \sum_{\sigma=\uparrow,\downarrow} O_{o,\sigma} (\bold{R}) \right)^2,
\label{int_rsp}
\end{equation}
where
\begin{align}
O_{o, \sigma} (\bold{R}) &= \frac{1}{3} Q_{o,\sigma} (\bold{R}) + \alpha_T T_{o,\sigma} (\bold{R}), \\
Q_{o,\sigma} (\bold{R}) &= \sum_{p=1}^6 n_{o\sigma p} (\vec R)\\
T_{o,\sigma} (\bold{R}) &= \sum_{p=1}^6 (-1)^{p-1}\left[ b^\dagger_{o\sigma p}(\vec R)+b_{o\sigma p}(\vec R) \right],
\end{align}
The sum runs over the centers of the honeycomb superlattice $\bold{R}$ and the electrons' spin $\sigma$ and valley $o$.  The $Q$ term sums over all electron densities $n_{o\sigma p}=c^\dagger_{o\sigma p}c_{o\sigma p}$ on the six sites $p=1\ldots6$ of the hexagon centered at $\vec R$, $T$ includes all nearest-neighbor hopping operators $b_{o\sigma p}=c^\dagger_{o\sigma p+1}c_{o\sigma p}$ along the hexagon.
The parameter $\alpha_T$ measures the relative strength of the non-local terms in (\ref{int_rsp}). Transforming $H_{int}$  to the band basis and projecting  it onto the patches around the Van Hove points, we obtain the bare coupling constants for the patch Hamiltonians (\ref{eq:iap6}),(\ref{eq:iap12})~\cite{Chichinadze2019nematic}. The two parameters $u$ and $g$ in the six-patch model and the five parameters $u, g_{1-}, g_{1+}, g_2, g_3$ in the twelve-patch model  are all proportional to $V_0$ and  are functions of $\alpha_T$. We find
\begin{equation}
u = V_0; \quad  g = V_0 \left(0.1 +0.92 \alpha^2_T\right)\, ,
\label{coupl_1}
\end{equation}
and
\begin{equation}
\begin{gathered}
u=V_0;      \quad g_2 = V_0 \left(0.193 + 0.053 \alpha^2_T \right) ; \quad g_{1-} = V_0 \left( 0.021 + 1.51 \alpha^2_T \right) ;  \\
 g_{1+}= V_0 \left( 0.256 + 17.9 \alpha^2_T \right) ; \quad  g_{3} = V_0 \left( 0.057 + 9.02 \alpha^2_T \right) .
 \end{gathered}
\label{coupl_2}
\end{equation}
Note that there are no terms linear in  $\alpha_T$. Such terms come from high-energy processes, which involve both valence and conduction bands, and therefore do not contribute to the low-energy theory.

\section{Density-wave and Pomeranchuk orders in the 6-patch model}
\label{sec:RPA6patch}

\begin{figure}[t]
\includegraphics[width=0.6\linewidth]{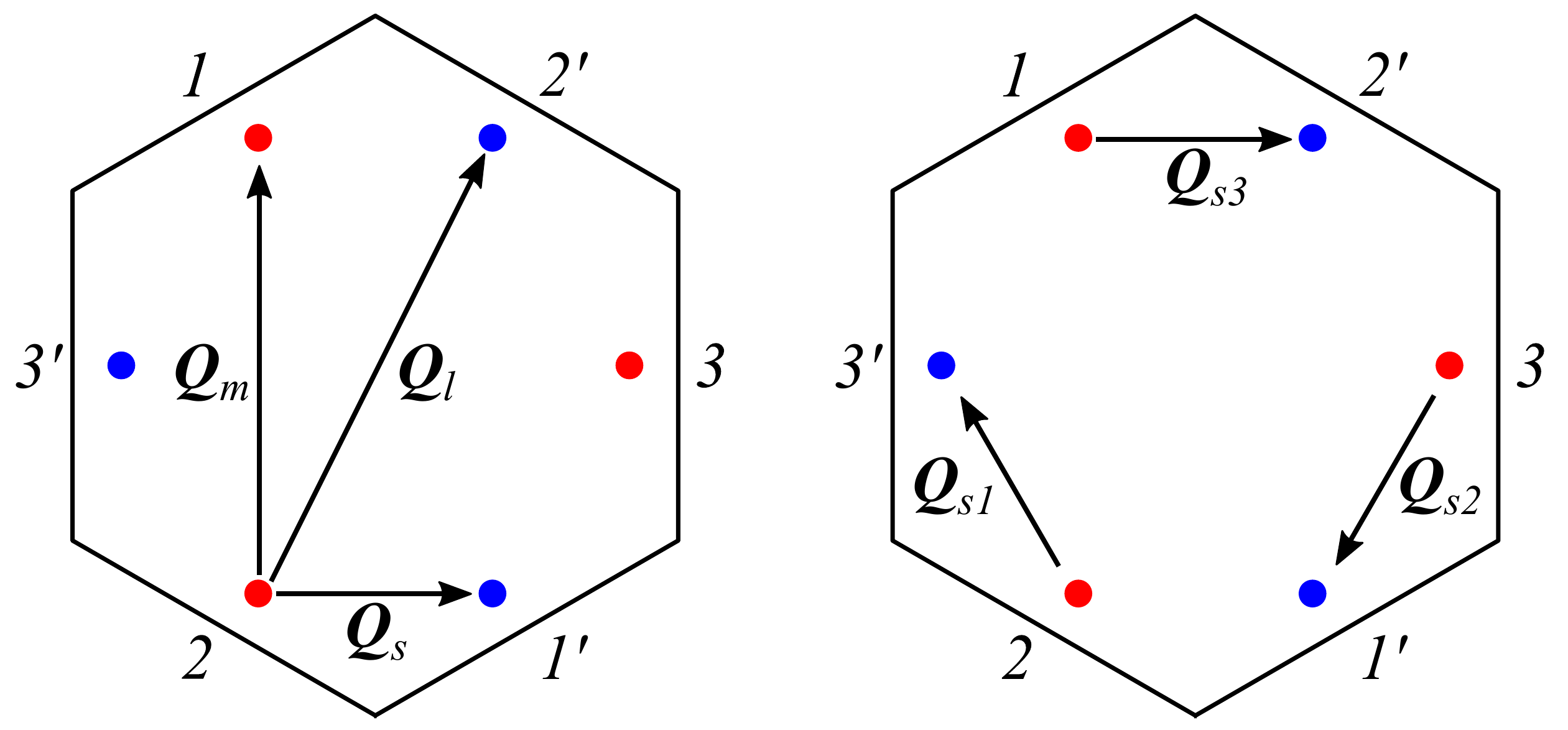}
\caption{
Momentum transfers between patches in the six-patch model. Blue and red dots mark the Van Hove points made of different valleys. There are three different types of momentum transfers, ${Q}_s, {Q}_m$, and ${Q}_l$.
There are three non-equivalent vectors of each type (right panel).
 Momenta ${Q}_m$ connect patches within the same valley, 
 while momenta ${Q}_s$ and ${Q}_l$ connect patches from different valleys. 
 }
\label{mod_sk6p}
\end{figure}

In the strict weak coupling limit, the leading instability in any patch model is superconductivity if there exists a pairing channel with an attractive interaction because it has a divergence $\sim\ln^2 T$. However, if the pairing interaction is repulsive, or if the coupling is moderate, the leading instability may instead be in the particle-hole channel, which diverges like $\sim\ln T$ due to the singular density of states.
In our previous work \cite{Chichinadze2019nematic}, we analyzed the couplings in particle-particle channels and identified the ones where the attraction is the strongest.  Here, we obtain the couplings in particle-hole channels.  We consider SDW and CDW channels with  the three different momenta,  $\vec{Q}_s, \vec{Q}_m$, and $\vec{Q}_l$, connecting Van Hove points, see Fig. \ref{mod_sk6p},  and spin and charge Pomeranchuk channels with zero transferred momentum, but different form factors.
For each $|Q_i|$ $i \in \{s,m,l\}$, there are
three
nonequivalent vectors connecting different patches.

To obtain the couplings in different channels, we introduce infinitesimally small bare particle-hole vertices
$\Gamma^0_j(\vec Q)$ with  momentum transfers $\vec Q\in\{\vec 0,\vec Q_s,\vec Q_m,\vec Q_l\}$ and the structure of CDW, SDW, and charge or spin Pomeranchuk order parameters. The label $j\in$ \{CDW,SDW,CPom,SPom\}.
This gives eight different vertices, which we list  in  Table \ref{order_table}.
The bare particle-hole vertices receive corrections due to interactions, which we
calculate by summing up series of ladder diagrams.
In this study,  we do not include mixed diagrams, which couple  renormalizations in the particle-particle and particle-hole channels.

In a patch model, a vertex $\Gamma_j(\vec Q)$ with given $j$ and ${\vec Q}$ is  a vector,  with components in different patches, and the  dressed vertices are
 \beq
 \Gamma_j({\vec Q})=\Gamma_j^0({\vec Q}) +\Pi({\vec Q})\Lambda_{j, Q} \Gamma_j({\vec Q}),
 \label{ll}
  \eeq
  where $\Pi({\vec Q})$ is the polarization bubble at momentum ${\vec Q}$ and  $\Lambda_{j}$  (the matrices in patch space) contain the information about intra-patch and inter-patch interactions.
Diagonalizing the equations, we obtain
\beq
{\bar \Gamma}_{j}({\vec Q}) = \frac{{\bar \Gamma}^0_{j}({\vec Q})}{1 - \Pi({\vec Q})\lambda_{j \, Q}},
\eeq
The  eigenvectors ${\bar \Gamma}_{j}({\vec Q})$ are linear combinations of $\Gamma_{j}({\vec Q})$, and  $\lambda_{j \, Q}$ are the eigenvalues of the matrix equation (\ref{ll}).

Our goal is to determine $\Pi({\vec Q}_i)$ and $\lambda_{j \, Q,i}$ in different particle-hole channels, and identify the channels with the largest
 attractive interaction.
   Within mean-field approximation, an instability in one of these
  channels develops when
$\Pi(\vec{Q})\lambda_{j \, Q,i} =1$.
All polarization bubbles scale logarithmically with temperature due to singular behavior of the density of states, hence
 the leading channel
is the one in which the prefactor for  $\ln T$ is the largest.

For a magnetic order, which breaks $O(3)$ spin-rotational symmetry, mean-field instability temperature in 2D
 determines the onset for a rapid increase of the correlation length, while the actual long-range order does not
 develop down to $T=0$ by Mermin-Wagner theorem.  In TBG, there is some coupling in the third direction due to, e.g., the substrate, and the actual instability temperature is finite, although smaller than the mean-field one.

\begin{table}[h]
\centering
\caption{List of all possible bilinear combinations of low-energy fermions near Van Hove points,
  classified into order parameters in charge and spin Pomeranchuk and density-wave channels.
  The spin order parameters are vectors. }
\label{order_table}
\begin{adjustbox}{width=\columnwidth,center}
 \begin{tabular}{||c | c | c | c| c|c||}
 \hline
 Order & Vertex  & Patch order parameters & Fermionic bilinear & Number of fields & Real or complex \\  [1ex]
 \hline \hline
 Charge $\vec{Q}=0$ (Pom) & $\Gamma_c(0)$ & $\Delta^c_i (0) $ and $\Delta^c_{i'} (0) $ & $ \av{f^{\dagger}_{i\sigma}  f_{i\sigma}}$ and $ \av{f^{\dagger}_{i's}  f_{i's}}$ & 6 & Real \\
 \hline
 Spin $\vec{Q}=0$ (Pom) & $\Gamma_s(0)$& $\Delta^s_i (0) $ and $\Delta^s_{i'} (0) $ & $\av{f^{\dagger}_{i\sigma} \vec{\sigma}_{\sigma \sigma'} f_{i\sigma'}}$ and $\av{f^{\dagger}_{i'\sigma} \vec{\sigma}_{\sigma \sigma'} f_{i'\sigma'}}$ & 6 & Real \\
 \hline
 Charge $\vec{Q}_s$ & $\Gamma_c(\vec Q_s)$ & $\Delta^c_i (\vec{Q}_s) $ and $\Delta^c_{i'} (\vec{Q}_s) $ & $\av{f^{\dagger}_{(i+2)'\sigma}  f_{(i+1)\sigma}}$ and $\av{f^{\dagger}_{(i+1)'\sigma}  f_{(i+2)\sigma}}$ & 6 & Complex\\
 \hline
 Spin $\vec{Q}_s$ & $\Gamma_s(\vec Q_s)$ & $\Delta^s_i (\vec{Q}_s) $ and $\Delta^s_{i'} (\vec{Q}_s) $ & $\av{f^{\dagger}_{(i+2)'\sigma} \vec{\sigma}_{\sigma \sigma'} f_{(i+1)\sigma'}}$ and $\av{f^{\dagger}_{(i+1)'\sigma} \vec{\sigma}_{\sigma \sigma'} f_{(i+2)\sigma'}}$ & 6 & Complex\\
 \hline
 Charge $\vec{Q}_m$ & $\Gamma_c(\vec Q_m)$ & $\Delta^c_i (\vec{Q}_m) $ and $\Delta^c_{i'} (\vec{Q}_m) $ & $\av{f^{\dagger}_{(i+2)\sigma}  f_{(i+1)\sigma}}$ and $\av{f^{\dagger}_{(i+1)'\sigma}  f_{(i+2)'\sigma}}$ & 6 & Complex\\
 \hline
 Spin $\vec{Q}_m$ &$\Gamma_s(\vec Q_m)$ & $\Delta^s_i (\vec{Q}_m) $ and $\Delta^s_{i'} (\vec{Q}_m) $ & $\av{f^{\dagger}_{(i+2)\sigma} \vec{\sigma}_{\sigma \sigma'} f_{(i+1)\sigma'}}$ and $\av{f^{\dagger}_{(i+1)'\sigma} \vec{\sigma}_{\sigma \sigma'} f_{(i+2)'\sigma'}}$ & 6 & Complex\\
 \hline
 Charge $\vec{Q}_l$& $\Gamma_c(\vec Q_l)$ & $\Delta^c_i (\vec{Q}_l)
 ={\Delta^c_{i'}} (\vec{Q}_l)^\dagger
 $  & $\av{f^{\dagger}_{i'\sigma}  f_{i\sigma}}$ & 3 & Complex\\
 \hline
 Spin $\vec{Q}_l$ &$\Gamma_s(\vec Q_l)$ & $\Delta^s_i (\vec{Q}_l)
 ={\Delta^s_{i'}} (\vec{Q}_l)^\dagger
 $  & $\av{f^{\dagger}_{i'\sigma} \vec{\sigma}_{\sigma \sigma'} f_{i\sigma'}}$ & 3 & Complex\\
 \hline
\end{tabular}
\end{adjustbox}
\end{table}

\subsection{The polarization bubbles}

\begin{figure}[h]
\includegraphics[width=0.45\linewidth]{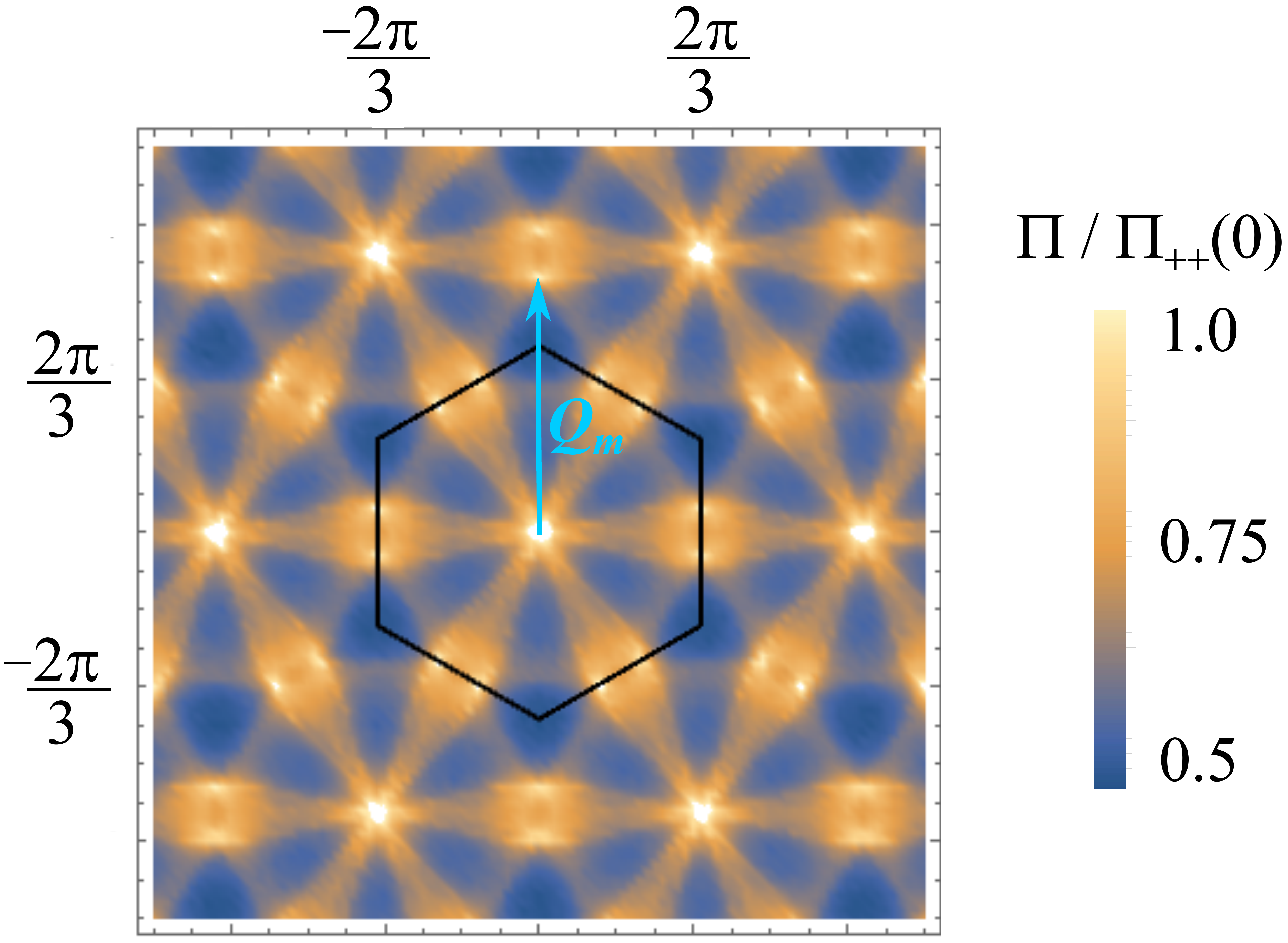} \quad
\includegraphics[width=0.45\linewidth]{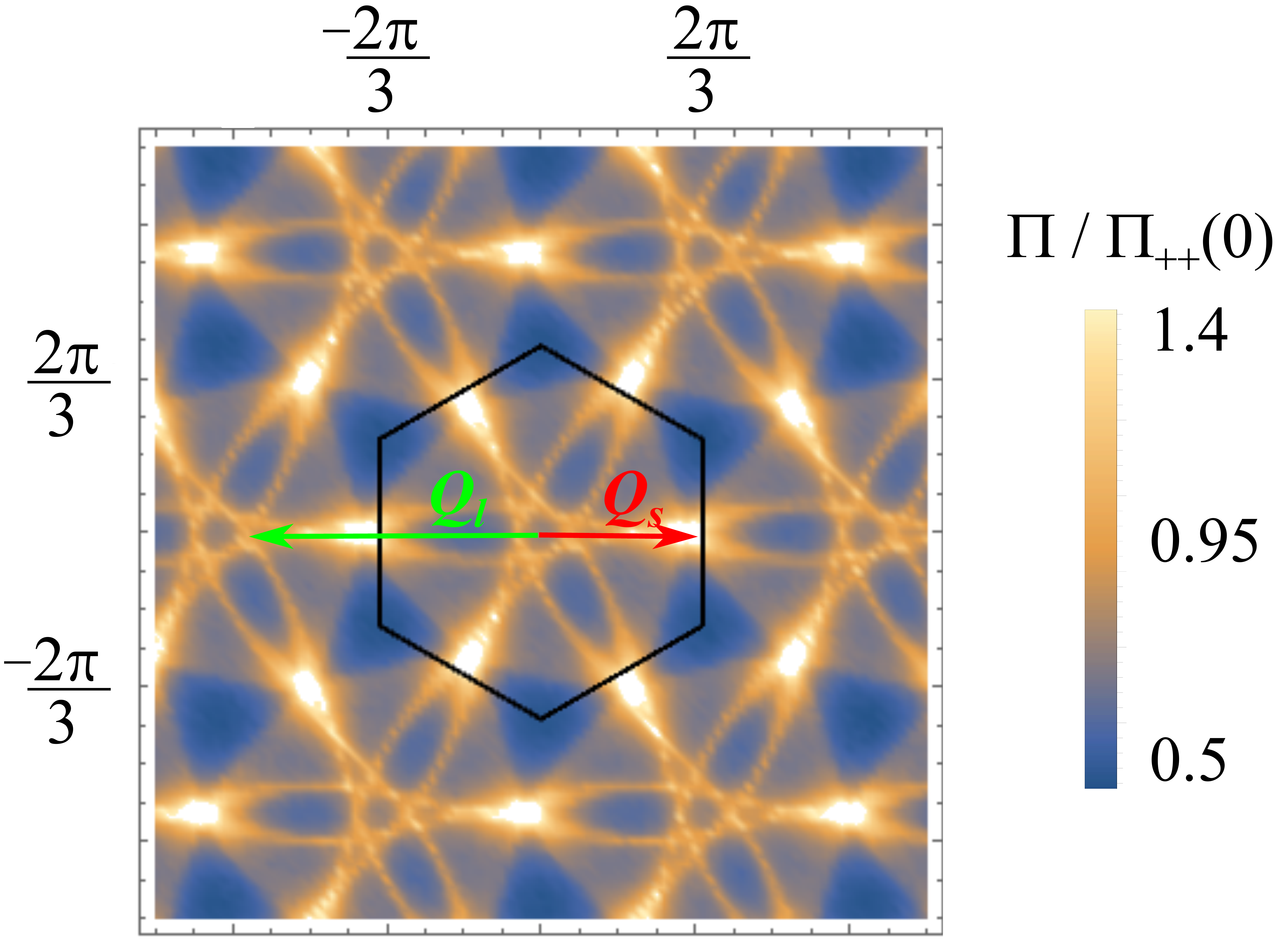}
\caption{ Plots of intra-valley polarization bubble $\Pi_{++}$ (left) and inter-valley polarization bubble $\Pi_{+-}$ (right), calculated for $T=0$. We moved the chemical potential away from the Van Hove doping by $\delta \mu \sim 0.001$,
to regularize the logarithmic divergence.
The black hexagon shows the Brillouine zone boundary. Color coding reflects the magnitude of the polarization bubble relative to $\Pi_{++} (0)$. Note that here the momenta $Q_m$, $Q_l$, and $Q_s$ all originate at the center of the Brillouine zone.}
\label{pic:polar6p}
\end{figure}

We first analyze the polarization bubbles. Explicitly, they are defined by
$\Pi_{op} (q)=-T \sum_\omega \int d\vec k\,G_o (k) G_p (k+q)>0$,  $o,p=\pm$ with the Green's functions $G_o(k)=1/(i\omega-E^{v,c}_o(k))$ (see Eq.~\ref{band_spec}). We consider the two bands that cross the Fermi surface, i.e. $E^{v}_\pm(k)$ ($E^{c}_\pm(k)$) for $\mu<0$ $(\mu>0)$, and distinguish the intra-valley $o=p$ and inter-valley $o\neq p$ polarization bubble. Due to time-reversal symmetry, we have $\Pi_{++}=\Pi_{--}$ and $\Pi_{+-}=\Pi_{-+}$ and due to rotation symmetry $\Pi_{op}(\vec Q_i)$ only depends on $|Q_i|$ and not on patch indices. Polarization bubbles at Van Hove doping are logarithmically divergent, i.e. $\Pi_{op}(\vec{Q}_i) \sim \ln \frac{\Xi}{\mathrm{max}(T, \mu)},$ where $\Xi$ is the UV cut-off.
We show the intra- and inter-valley polarization bubbles in Fig.~\ref{pic:polar6p}. For low enough temperatures or for $T=0$ but small offset from the Van Hove doping, the peaks at zero and the different momenta $\vec Q_i$ are clearly visible. We find that $\Pi_{+-}(\vec Q_s)\gtrsim\Pi_{++}(\vec 0)\gtrsim\Pi_{++}(\vec Q_m)\gtrsim\Pi_{+-}(\vec Q_l)$. We give exemplary values in Tab.~\ref{polar_table}. This hierarchy remains qualitatively the same if we vary the microscopic hopping parameters. The reason is that the degree of approximate nesting is larger for $\vec Q_s$ than $\vec Q_m$ and $\vec Q_l$. for  For $t_3\rightarrow 0$, the differences become smaller because the Van Hove points move closer to the Brillouin zone boundary, i.e. $|\vec Q_s|,|\vec Q_m|$ and $|\vec Q_l|$ approach each other.

\begin{table}[h!]
\centering
\caption{Intra-valley $\Pi_{++}$ and inter-valley $\Pi_{+-}$ polarization bubbles, normalized to $\Pi_{++} (0)$, for the six-patch model at $T=0$. Like before, we moved the chemical potential by $\delta \mu \sim 0.001$ away from the Van Hove doping to regularize logarithmic divergencies.
 $G_\pm(k)=1/(i\omega-E_\pm(k))$ are the Green's functions of fermions from different valleys.}

\label{polar_table}
 \begin{tabular}{||c | c | c||}
 \hline
 Polarization operator $\Pi_{op} (\vec{Q})$ & Green's functions & $\Pi_{op} (\vec{Q}) / \Pi_{++} (0)$  \\  [1ex]
 \hline \hline
 $\Pi_{++} (0)$ & $-\int G_+(k) G_+(k)$ & 1 \\
 \hline
 $\Pi_{+-} (\vec{Q}_s)$ & $-\int G_+(k) G_-(k+Q_s)$ & 1.4  \\
 \hline
 $\Pi_{++} (\vec{Q}_m)$ & $-\int G_+(k) G_+(k+Q_m)$ &  0.96  \\
 \hline
 $\Pi_{+-} (\vec{Q}_l)$ & $-\int G_+(k) G_-(k+Q_l)$ & 0.84 \\
 \hline
\end{tabular}
\end{table}

\subsection{The dressed vertices}

\begin{figure}[t]
\center{\includegraphics[width=0.99\linewidth]{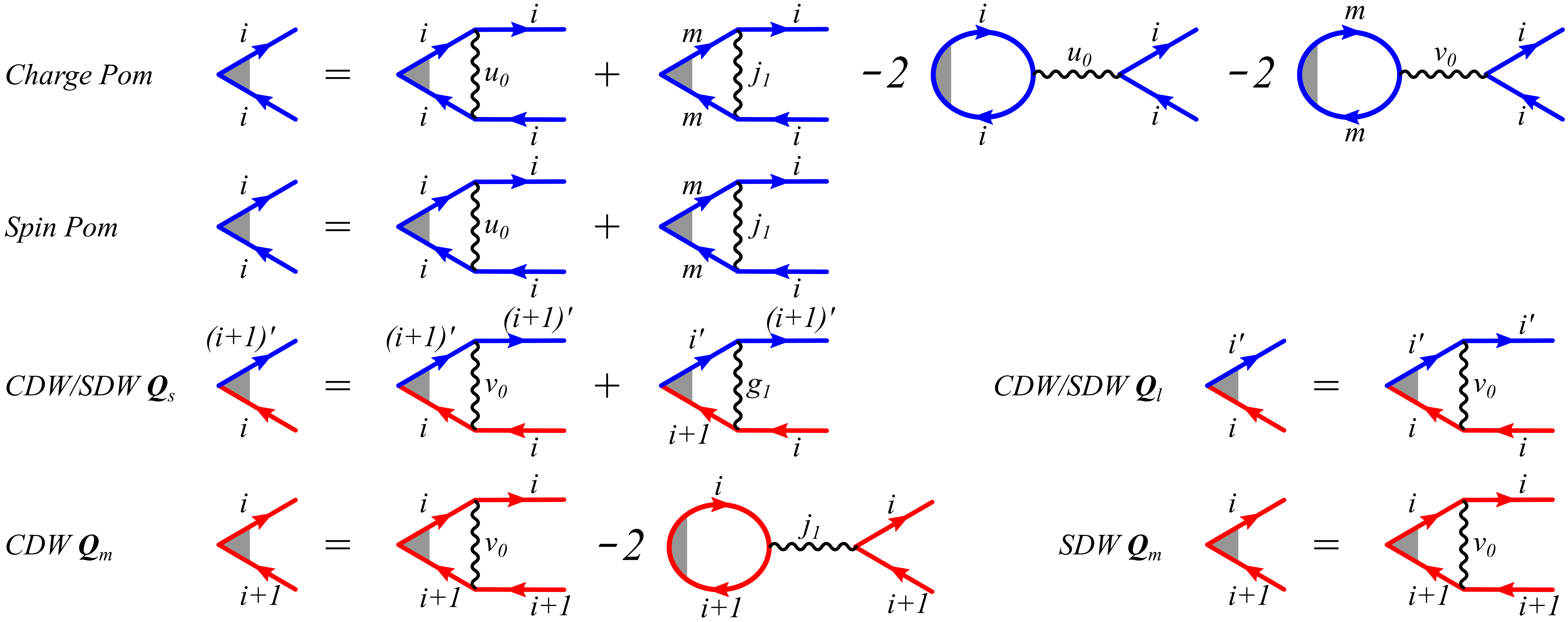} }
\centering{}\caption{Diagrammatic representation of a system of coupled equations for dressed vertices. Gray triangle is a fully renormalized vertex, red and blue lines are Green's functions of electrons from the two valleys. Summation over $m$ is implied. When a diagrammatic equation involves fermions of only one color, there is an identical equation for fermions of the other color.
 The bare vertices are not shown for shortness.
}
\label{gapeq_pic}
\end{figure}

Next, we introduce trial vertices in different ordering channels, dress them up by interactions, and discuss the structure of the dressed vertices.
  We show the diagrammatic expressions for the dressed vertices in Fig. \ref{gapeq_pic}. In the Pomeranchuk channel, the order parameters are bilinears in fermionic operators from the same patch and the same valley with zero momentum transfer. They can be in either the spin or the charge channel. We introduce $\Gamma_c(0)=[\Delta^c_1 (0),\Delta^c_2 (0),\Delta^c_3 (0),\Delta^c_{1'} (0),\Delta^c_{2'} (0),\Delta^c_{3'} (0)]$ and $\Gamma_s(0)=[\Delta^s_1 (0),\Delta^s_2 (0),\Delta^s_3 (0),\Delta^s_{1'} (0),\Delta^s_{2'} (0),\Delta^s_{3'} (0)]$, where $\Delta_{i^{(\prime)}}^{c}(\Delta_{i^{(\prime)}}^{s})$ is the dressed vertex for charge (spin) Pomeranchuk order at patch $i^{(\prime)}$ (see Tab.~\ref{order_table}). The Pomeranchuk vertices describe intra-valley, intra-patch ordering tendencies. The ladder series for the dressed vertices yields (see Fig.~\ref{gapeq_pic})
\begin{align}
\Gamma_{CPom}(0)&=\Gamma_{CPom}^0(0)+\Pi_{++}(0)\Lambda_{CPom,0}\Gamma_{CPom}(0)\\
\Gamma_{SPom}(0)&=\Gamma_{SPom}^0(0)+\Pi_{++}(0)\Lambda_{SPom,0}\Gamma_{SPom}(0)
\end{align}
with
\beq
\Lambda_{CPom,0}=
\begin{pmatrix}
-u & g-2u & g-2u & 0 & 0 & 0 \\
g-2u & -u & g-2u & 0 & 0 & 0 \\
g-2u & g-2u & -u & 0 & 0 & 0 \\
0 & 0 & 0  & -u & g-2u & g-2u \\
0 & 0 & 0 & g-2u & -u & g-2u \\
0 & 0 & 0 & g-2u & g-2u & -u
\end{pmatrix}
\qquad
\Lambda_{SPom,0}=
\begin{pmatrix}
u & g & g & 0 & 0 & 0 \\
g & u & g & 0 & 0 & 0 \\
g & g & u & 0 & 0 & 0 \\
0 & 0 & 0 & u & g & g \\
0 & 0 & 0 & g & u & g \\
0 & 0 & 0 & g & g & u
\end{pmatrix}
\eeq
We see that the two components from different valleys $[\Delta^{c(s)}_1 (0),\Delta^{c(s)}_2 (0),\Delta^{c(s)}_3 (0)]$ and $[\Delta^{c(s)}_{1'} (0),\Delta^{c(s)}_{2'} (0),\Delta^{c(s)}_{3'} (0)]$ decouple, i.e.
there are two independent series of ladder renormalizations for fermions from different valleys.

For SDW and CDW channels, the fermionic bilinears are formed by an electron and a hole from different patches, and from the same valley (the order with momenta $\vec Q_m$) or from opposite valleys (the orders with momenta $\vec Q_s,\vec Q_l$), see Fig.~\ref{mod_sk6p}.
We introduce the charge and spin vertices $\Gamma_{c,s}(\vec Q)=[\Delta^{c,s}_1 (\vec{Q}),\Delta^{c,s}_2 (\vec{Q}),\Delta^{c,s}_3 (\vec{Q}),\Delta^{c,s}_{1'} (\vec{Q}),\Delta^{c,s}_{2'} (\vec{Q}),\Delta^{c,s}_{3'} (\vec{Q})]$ with order parameters $\Delta^{c,s}_i (\vec{Q})$ connecting the different patches and characteristic momentum transfer $\vec Q\in\{\vec Q_s, \vec Q_m,\vec Q_l\}$ (see Tab.~\ref{order_table}).
 The dressed vertices for CDW and SDW are of the general form
\begin{align}
\Gamma_{CDW}(\vec Q)=\Gamma_{CDW}^0(\vec Q)  +\Pi_{+-}(\vec Q)\Lambda_{CDW \, Q}\Gamma_{CDW}(\vec Q) \\
\Gamma_{SDW}(\vec Q)=\Gamma_{SDW}^0(\vec Q)  +\Pi_{+-}(\vec Q)\Lambda_{{SDW}\, Q}\Gamma_{SDW}(\vec Q)
\end{align}
If $\vec Q=\vec Q_m$, the polarization bubble is intra-valley $o=p$ and the coupling matrix is diagonal
\begin{align}
\Lambda_{CDW}(\vec Q_m)&=(u-2g)\mathbbm{1}\\
\Lambda_{SDW}(\vec Q_m)&=u\mathbbm{1}\,.
\end{align}
If $\vec Q=\vec Q_s$ or  $\vec Q=\vec Q_l$, the polarization bubble is inter-valley $o\neq p$. In this case, the ladder series for SDW and CDW are formed by the same type of diagrams, because the diagrams that usually distinguish charge and spin channels are absent when valley mixing is not allowed (see Fig.~\ref{gapeq_pic}).
That means in the equations for $\Gamma_c(\vec Q_s)$ and $\Gamma_s(\vec Q_s)$
the coupling matrices for spin and charge channel are the same
\beq
\Lambda_{SDW }(\vec Q_s)=\Lambda_{SDW }(\vec Q_s)=\begin{pmatrix}
u & 0 & 0 & g &0 &0 \\
0 & u & 0  &0 & g &0\\
0 & 0 & u & 0 &0 & g \\
g & 0& 0 &u & 0&0 \\
0& g & 0 &0 &u & 0 \\
0& 0& g & 0& 0 & u
\end{pmatrix}
\eeq
and
\beq
\Lambda_{CDW}(\vec Q_l)=\Lambda_{SDW }(\vec Q_l)=u \mathbbm{1}\,.
\eeq

\begin{figure}[t]
\center{\includegraphics[width=0.99\linewidth]{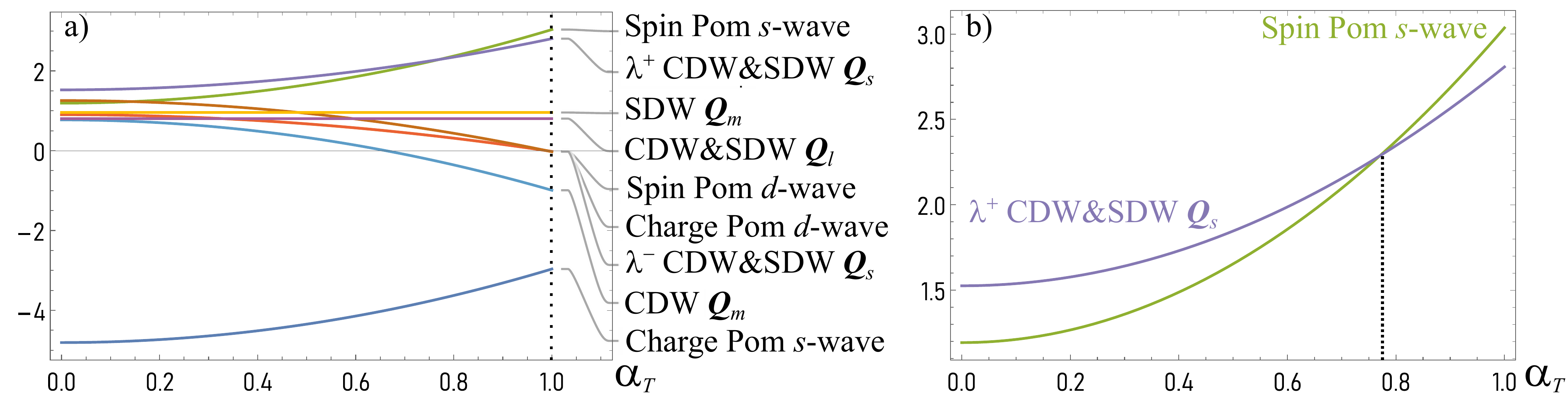} }
\centering{}\caption{The eigenvalues (the products of the interactions and the polarization bubbles) for the six-patch model as functions of $\alpha_T$.
 We used normalized polarization bubbles from Table \ref{polar_table} to avoid the logarithmic factor.
  A positive eigenvalue means an attraction in the corresponding ordering channel.
  Panel a): The eigenvalues in all channels. The dashed line shows the onset of superconductivity.
   Panel b): The two channels with the strongest attractive eigenvalues --  SDW/CDW with $\vec{Q}_s$ and $s-$wave spin Pomeranchuk with $Q=0$. In this panel the dashed line indicates the degeneracy point of two eigenvalues.  }
\label{lambda_alpha}
\end{figure}

Observe that the matrices   $\Lambda_{j \, Q}$ are either block diagonal, or can be made block-diagonal by a simple permutation of the order parameters. Therefore, every eigenvalue  is $N$ times degenerate, where $N$ -- is the number of identical blocks in the matrix.
Diagonalizing the blocks, we obtain the eigenvalues and eigenvectors for every channel. The eigenvalues coincide with the couplings of the channel and the eigenvectors encode the corresponding symmetry.  Overall, we find nine different eigenvalues: four in the Pomeranchuk channels and five in SDW/CDW channels.
In the charge Pomeranchuk channel, we find
\begin{align}
\lambda^{s}_{CPom}& =\Pi_{++}(0) \left(2 g- 5u\right) \\
\lambda^{d}_{CPom} &=\Pi_{++}(0) \left(u  -g\right)\,.
\label{align}
\end{align}
For $\lambda^{s}_{CPom}$,  the eigenvector is $(1,1,1)$, so it is natural to call this state $s-$wave.  The eigenvalue $\lambda^{d}_{CPom}$ is doubly degenerate with the two eigenvectors $(0,1,-1)$ and $(1,-1/2,-1/2)$. This state is often called $d-$wave
because of its symmetry.
 The same situation holds for the spin Pomeranchuk channel.  Here
\begin{align}
\lambda_{SPom}^{s} &= \Pi_{++}(0) \left(u + 2 g \right)\\
\lambda_{SPom}^{d} &= \Pi_{++}(0) \left( u- g \right)\,.
\label{eq_FM}
\end{align}
For CDW and SDW orders with $\vec{Q}_m$, the coupling matrices of the ladder series are diagonal, thus the eigenvectors are trivial and the eigenvalues can be read off
\begin{align}
\lambda_{CDW}(\vec Q_m) &= \Pi_{++}(\vec{Q}_m) \left(u - 2 g\right)\\
\lambda_{SDW}(\vec Q_m) &= \Pi_{++}(\vec{Q}_m) u\,.
\label{eq_Qm}
\end{align}
For density wave orders with $\vec{Q}_s$ the situation is different. This time the number of identical blocs is $N=3$,  hence every eigenvalue of a block is triply degenerate.
The blocks are $2 \times 2$ matrices involving fields $\Delta^{c,s}_1 (\vec{Q}_s)$ and $\Delta^{c,s}_{1'} (\vec{Q}_s)$, etc, hence every block corresponds to one of momentum transfer vectors $\vec{Q}_{s1}, \vec{Q}_{s2},$ and $\vec{Q}_{s3}$ (see Fig.~
\ref{mod_sk6p}).
The eigenvalues now read
\begin{align}
\lambda^{+}_{CDW} (\vec Q_s)&=\lambda^{+}_{SDW}(\vec Q_s)=\Pi_{+-}(\vec{Q}_s) \left(u + g \right) \\
\lambda^{-}_{CDW}(\vec Q_s) &=\lambda^{-}_{SDW}(\vec Q_s) =\Pi_{+-}(\vec{Q}_s) \left(u - g \right)
\label{eq_Qs}
\end{align}
with superscript $+/-$ corresponding to eigenvectors $(1,\pm 1)$ for every block.
The coupling matrix for CDW and SDW with $\vec Q_l$ is again diagonal, and the eigenvalue is given by
\begin{equation}
\lambda_{CDW}(\vec Q_l) =\lambda_{SDW}(\vec Q_l)=\Pi_{+-}(\vec{Q}_l) u\,.
\label{eq_Ql}
\end{equation}

\subsection{The eigenvalues}

We can now compare the eigenvalues in the different channels to determine the one with the largest critical temperature for varying $\alpha_T$.
To this end, we use Eq.~(\ref{coupl_1}) for the interactions and Table \ref{polar_table} for the polarization bubbles. We show the eigenvalues as functions of $\alpha_T$ in Fig. \ref{lambda_alpha}.

We see that in several channels the eigenvalues are attractive even for $\alpha_T = 0$
(see Fig. \ref{lambda_alpha}). This is the consequence of the cluster nature of the Hubbard-like term in the
 microscopic model of Eq.~\eqref{int_rsp}.
If the interaction was purely local, the only positive eigenvalue would be in
the  $s-$wave spin Pomeranchuk (FM) channel.  The cluster Hubbard-like term contains non only on-site interaction, but also interactions  between fermionic densities at different sites of a particular hexagon.
This effectively introduces non-locality  and generates positive (attractive) eigenvalues  in some channels.
 For $\alpha_T\neq0$,   there is an additional momentum dependence from the pair-hopping
 and exchange-like
 interaction terms.

 We find that the two largest eigenvalues are in the degenerate CDW and SDW channel with momentum $\vec Q_s$
    and in the $s$-wave spin-Pomeranchuk channel (an intra-valley ferromagnetic channel).    For $\alpha_T \lesssim 0.77$, the eigenvalue in CDW/SDW channel is larger,  which we can be traced back to the fact that $\Pi_{+-}(\vec Q_s)$ is the largest polarization bubble. However,  for $\alpha_T\gtrsim 0.77$, the  eigenvalue in the $s$-wave spin-Pomeranchuk channel becomes the largest.
 The eigenvalues in some other channels are also attractive, but are smaller. However, the magnitudes of the eigenvalues can be  affected by, e.g., the coupling between renormalizations in the particle-particle and particle-hole channels (this effect is captured within, e.g., parquet and functional RG).
    In particular, an attraction in a $d$-wave Pomeranchuk channel can potentially become the strongest, as it was argued to happen in other systems~\cite{2020arXiv200614729C}.
    We  argue in Sec.~\ref{sec:dPoms} below that, if this happens, lattice-rotational symmetry gets spontaneously broken, i.e., the ground state becomes a nematic.

\section{Landau functional for the six-patch model}
\label{sec:Landau6patch}

\begin{figure}[t]
\center{\includegraphics[height=0.15\linewidth]{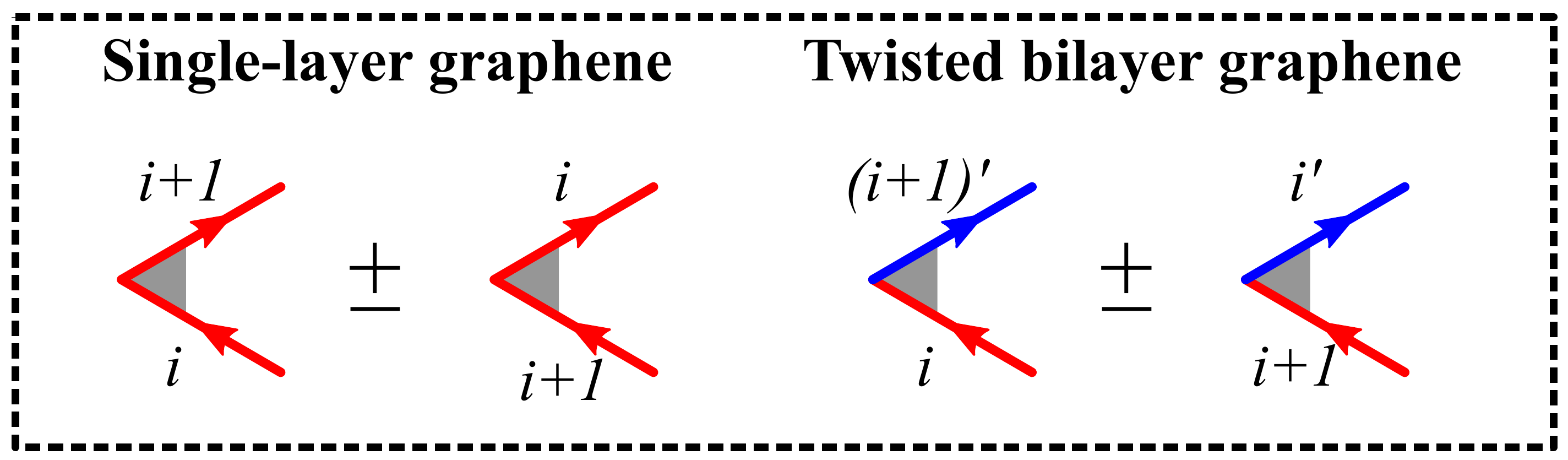} }
\centering{}\caption{The linear combinations of particle-hole order parameters in twisted bilayer graphene and in single-layer graphene, each at Van Hove doping.   In single-layer graphene, low-energy excitations involve only one type of fermions,  and the linear combinations of the two conjugated vertices are either purely real or purely imaginary. In twisted bilayer graphene, the two fermions in a particle-hole vertex are from different valleys (bands), and the linear combinations of the two conjugated vertices are neither purely real nor purely imaginary.}
\label{comb_sketch}
\end{figure}

In this section we derive the Landau free energy for different order parameters. This
 will allow
  us to determine the structure of the ordered state. We cannot determine this structure at the quadratic level because each leading eigenvalue is degenerate.

\subsection{Order parameters}
 For SDW/CDW, the order parameters
 with the largest eigenvalues
 are symmetric combinations of
 ${\bf \Delta}_i^s ({\bf Q}_s) = \av{f^{\dagger}_{(i+2)'\sigma} \vec{\sigma}_{\sigma \sigma'} f_{(i+1)\sigma'}}$ and
  ${\bf \Delta}_{i'}^s ({\bf Q}_s) = \av{f^{\dagger}_{(i+1)' \sigma} \vec{\sigma}_{\sigma \sigma'} f_{(i+2) \sigma'}}$ in the spin channel and of ${ \Delta}_i^c ({\bf Q}_s) = \av{f^{\dagger}_{(i+2)'\sigma} f_{(i+1)\sigma}}$ and
  ${ \Delta}_{i'}^c ({\bf Q}_s) = \av{f^{\dagger}_{(i+1)' \sigma} f_{(i+2)\sigma}}$ in the charge channel.
Accordingly, we introduce
three
scalar fields $\Delta_i$  and three
vector fields ${\bf M}_i$ as
\begin{equation}
\begin{gathered}
\Delta_i = \Delta_i^c (\vec{Q}_s)+ \Delta_{i'}^c (\vec{Q}_s), \\
{\bf M}_i = {\bf \Delta}_i^s (\vec{Q}_s)+ {\bf\Delta}_{i'}^s (\vec{Q}_s),
\end{gathered}
\label{n_1}
\end{equation}
 We show the three vectors ${\bf Q}_{si}$ in Fig.~\ref{mod_sk6p}.
 Note that we have $\Delta_i^{s(c)}(-\vec Q_s)=\Delta^{s(c)}_{i'}(\vec Q_s)$, but not the complex conjugate $\Delta_i^{s(c)}(-\vec Q_s)\neq \bar \Delta_{i}^{s(c)}(\vec Q_s)$. Thus,
each
 ${\bf \Delta}_i $
 and each $\vec M_i$
 is a complex order parameter, because low-energy excitations at Van Hove points $i$ and $i'$ belong to different valleys (bands). In this respect,  our case is different from single-layer graphene, where all six Van Hove points come from the same band, and the dispersions at $i$ and $i'$ are identical.  In that case, $\Delta_i$ and ${\bf M}_i$ are real fields.  We illustrate this difference in Fig. \ref{comb_sketch}.

An $s-$wave spin Pomeranchuk order within a given valley  is equivalent to intra-valley ferromagnetism, and we found that the order on different valleys is decoupled. Accordingly, we introduce two three-component vector fields for ferromagnetic order within each valley sector
 \begin{align}
{\bf S} = \frac{1}{3}\sum_i {\bf \Delta}_i^s (0)\qquad
{\bf S}' = \frac{1}{3}\sum_{i'} {\bf \Delta}_{i'}^s (0)\,,
\end{align}
where ${\bf \Delta}_{i}^s (0) = \av{f^{\dagger}_{is} \vec{\sigma}_{ss'} f_{is'}}$ is the spin order parameter for a given patch (see Tab.~\ref{order_table}), and $\vec S$ and $\vec S'$ represent the total magnetization for each valley.
Because $d$-wave spin Pomeranchuk components are assumed to be zero,  ${\bf \Delta}_i^s$ and
${\bf \Delta}_{i'}^s$ are actually independent on $i$ for this order.

Finally, we introduce order parameters for $d$-wave charge and spin Pomeranchuk order.
Using that the two eigenvectors with $d$-wave symmetry are proportional to $(0,1,-1)$ and $(1,-1/2,-1/2)$ in each valley sector, we
  define
\beq
 \begin{aligned}
{\chi_{d1}} &=  \frac{1}{\sqrt{6}}[{2\Delta}_1^c (0)-\Delta_2^c (0)-\Delta_3^c (0)]\\
{\chi_{d2}} &=\frac{1}{\sqrt{2}}[{\Delta}_1^c (0)-\Delta_2^c (0)]
 \end{aligned}
 \qquad
  \begin{aligned}
{\chi'_{d1}} &= \frac{1}{\sqrt{6}}[{2\Delta}_{1'}^c (0)-\Delta_{2'}^c (0)-\Delta_{3'}^c (0)] \\
{\chi'_{d2}} &=\frac{1}{\sqrt{2}}[{\Delta}_{1'}^c (0)-\Delta_{2'}^c (0)]
\end{aligned}
\label{eq:opCPom}
\eeq
in the charge sector and accordingly in the spin sector
\beq
 \begin{aligned}
{\vec\phi_{d1}} &= \frac{1}{\sqrt{6}}[2{\bf\Delta}_1^s (0)-{\bf\Delta}_2^s (0)-{\bf\Delta}_3^s (0)]\\
{\vec\phi_{d2}} &= \frac{1}{\sqrt{2}}[{\bf\Delta}_1^s (0)-{\bf\Delta}_2^s (0)]
 \end{aligned}
 \qquad
  \begin{aligned}
{\vec\phi'_{d1}} &=\frac{1}{\sqrt{6}}[2{\bf\Delta}_{1'}^s (0)-{\bf\Delta}_{2'}^s (0)-{\bf\Delta}_{3'}^s (0)]\\
{\vec\phi'_{d2}} &=  \frac{1}{\sqrt{2}}[{\bf\Delta}_{1'}^s (0)-{\bf\Delta}_{2'}^s (0)]\,.
\end{aligned}
\label{eq:opSPom}
\eeq

\subsection{Quadratic free energy}

 For SDW/CDW with $\vec{Q}_s$, we have three independent complex scalar fields $\Delta_i$ and three independent complex vector fields ${\bf M}_i$.
 In addition, each field possesses an O(2) symmetry related to translational symmetry because the characteristic momentum transfer $\vec Q_s$ is incommensurate with the lattice.
 The quadratic part of the Landau functional can be deduced from the ladder renormalizations:
 \beq
 \label{eq:quadratic_CDW/SDW}
\mathcal{F}_{DW}^{(2)} \propto (1 - \lambda^+_{CDW/SDW}) \sum_i \left({\bar \Delta}_i \Delta_i + {\bar {\bf M}}_i {\bf M}_i \right)
 \eeq
 At this level, the order parameter manifold is huge: (U(1))$^6\times$
 (O(2))$^6  \times$(O(3))$^3$.

For the $Q=0$ Pomeranchuk channel, we have two ferromagnetic fields ${\bf S}$ and ${\bf S}'$.
 The quadratic part of the Landau functional is
\beq
 \label{eq:quadratic_FM}
\mathcal{F}_{S}^{(2)} \propto (1 - \lambda^s_{SPom}) \left[{\bf S}^2 +({\bf S}')^2\right]
 \eeq
  We see that it depends only on the sum of the squares of the order parameters, i.e., a  relative
   magnitudes of $|{\bf S}|$ and $|{\bf S}'|$ and a relative angle between ${\bf S}$ and ${\bf S}'$ are undetermined (the
   order parameter manifold  at this level is O(3)$\times$O(3)).

The charge and spin $d$-wave Pomeranchuk channel have the same eigenvalue, which itself is twofold degenerate. In addition, the valley sectors are decoupled. So, the quadratic part of the free energy is
\beq
 \label{eq:quadratic_dPOM}
\mathcal{F}^{(2)}_d\propto (1 - \lambda^d_{CPom/SPom}) \left[ \chi_{d1}^2 +\chi_{d2}^2+(\chi'_{d1})^2 +(\chi'_{d2})^2+\vec\phi_{d1}^2 +\vec\phi_{d2}^2+(\vec\phi'_{d1})^2 +(\vec\phi'_{d2})^2\right]\,.
 \eeq
The actual order is determined by terms beyond the quadratic level, which can substantially reduce the order parameter manifold.  We show the details of the derivation of the free energy to fourth order in the order parameter fields in Appendix~\ref{app:HStechs}.

\subsection{SDW/CDW ground state}

We first consider SDW/CDW
 order.
The total free energy consists of three terms: individual free energies for the CDW and SDW and a mixed term
 \begin{equation}
\mathcal{F}_{DW}=\mathcal{F}_{c}+\mathcal{F}_{s}+\mathcal{F}_{cs}
\end{equation}
with
\begin{align}
\label{eq:Fc}
\mathcal{F}_{c}&=
 \alpha\sum_i |\Delta_i|^2 + 2 Z_1 \sum_i|\Delta_i|^4+ 2Z_2 \sum_{i\neq j} |\Delta_i|^2 |\Delta_j|^2\\
\mathcal{F}_{s}&=
 \alpha\sum_i  \bar{\vec{M}}_i \cdot \vec{M}_i  +
2 Z_1 \sum \left(  2 (\bar{\vec{M}}_i \cdot \vec{M}_i)^2 - (\bar{\vec{M}}_i \cdot \bar{\vec{M}}_i) (\vec{M}_i \cdot \vec{M}_i) \right) \nonumber \\
&+ 2 Z_2 \sum_{i\neq j}\Big[ (\bar{\vec{M}}_i \cdot \vec{M}_j) (\vec{M}_i \cdot \bar{\vec{M}}_j) -(\bar{\vec{M}}_i \cdot \bar{\vec{M}}_j) (\vec{M}_i \cdot \vec{M}_j)   + (\bar{\vec{M}}_i \cdot\vec M_i) (\bar{\vec{M}}_j  \cdot\vec{M}_j)\Big]
 \label{eq:Fs}
\end{align}
and
\begin{align}
\mathcal{F}_{cs} &=
 8 Z_1 \sum_i \bar{\Delta}_i  \Delta_i  (\bar{\vec{M}}_i \cdot \vec{M}_i)
+ 2 Z_1 \sum_i \left[ \bar{\Delta}_i^2  (\vec{M}_i \cdot \vec{M}_i) +  \Delta_i^2  (\bar{\vec{M}}_i \cdot \bar{\vec{M}}_i) \right] \nonumber \\
&+ 2 Z_2\sum_{i\neq j} \Big[ \bar{\Delta}_i \bar{\Delta}_j  (\vec{M}_i \cdot \vec{M}_j)  + \Delta_i \Delta_j  (\bar{\vec{M}}_i \cdot \bar{\vec{M}}_j) + 2 \bar{\Delta}_i  \Delta_j  (\bar{\vec{M}}_j \cdot \vec{M}_i) + 2 \bar{\Delta}_i \Delta_i   (\bar{\vec{M}}_j \cdot \vec{M}_j)\Big],
\end{align}
where $\alpha\propto (1 - \lambda^+_{CDW/SDW})$ and
 $Z_1$ and $Z_2$
  are the convolutions of
  four fermionic Green's functions
$Z_1 = T\sum_\omega\int d\vec k G_o^2(\vec k) G_p(\vec k+\vec Q_{si})^2$, $Z_2 = T\sum_\omega \int d\vec k G_o^2(\vec k) G_p(\vec k+\vec Q_{si}) G_p(\vec k+\vec Q_{sj})$
 with  patch indices $i\neq j$ and
 opposite valley indices, i.e.,  if $o$ is $+$ then $p$ is $-$, and vice versa.
These
 $Z_1$ and $Z_2$ are independent of the patch indices due to rotation symmetry.
 At $T \to 0$, $\lambda^+_{CDW/SDW}$ diverges logarithmically, and $Z_1$ and $Z_2$ diverge as $1/T^2$,
  indicating that at a Van Hove filling there is no regular Landau expansion at $T=0$. We, however, are interested in the system behavior at a finite $T$, near a temperature for which $\lambda^+_{CDW/SDW} =1$.  For  a finite $T$, $Z_1$ and $Z_2$ are finite, and the Landau expansion is regular.
  We verified numerically that $Z_1\gg Z_2 > 0$.
  This indicates that the  transition is second order.

As a first step, we analyze
separately  $\mathcal{F}_c$ and $\mathcal{F}_s$.
In $\mathcal{F}_c$, the first quartic term sets the overall magnitude of $\sum_i\left| \Delta_i\right|$, while the second quartic term distinguishes the three different transfer vectors $Q_{si}$ (Fig.~\ref{comb_sketch}). Because $Z_2-Z_1<0$, $\mathcal{F}_c$ is minimized for $\left|\Delta_1\right|=\left|\Delta_2\right|=\left|\Delta_3\right|=\Delta \neq 0$. The relative phase between the fields $\Delta_i$ remains undetermined in Eq.~\eqref{eq:Fc}. In principle, an additional quartic term is allowed by symmetry $\delta\mathcal{F}_c^{(4)}\propto\sum_{i\neq j}\left(\bar\Delta_i^2\Delta_j^2 + \text{c.c.}\right)$, which would fix the phase. It involves fermions away from the patches so that its prefactor is suppressed and it does not appear in the patch approximation.
The prefactor was estimated to be negative \cite{Nandkishore2012PRL} in graphene (the calculation is analogous in our case), which favors the relative phase between the three $\Delta_{i}$ to be zero.

To analyze $\mathcal{F}_s$, we parameterize the fields by $\vec M_i=\exp(i\varphi_i)\vec m_i$ with real vector field $\vec{m}_i$. This leads to the free energy
\begin{align}
\mathcal{F}_{s}&=
 \alpha
  \sum_i  \vec{m}_i^2 + 2 Z_1  \left(   \vec{m}_1^2 + \vec{m}_2^2 + \vec{m}_3^2  \right)^2 + 4 (Z_2-Z_1) \left(\vec m_1^2 \vec m_2^2 + \vec m_1^2 \vec m_3^2 + \vec m_2^2 \vec m_3^2\right)\,.
\end{align}
Following the same reasoning as in the CDW case, we again find that a state with $\vec m_1^2=\vec m_2^2=\vec m_3^2=M^2$ minimizes the free energy, with undetermined angle and relative phase between the vectors $\vec m_i$. In  $\mathcal{F}_s$, the O(3)$\times$U(1) symmetry also permits a term $\delta\mathcal{F}_s^{(4)}\propto \sum_{i\neq j}\left[ \left(\bar{\vec M}_i \cdot {\vec M}_j\right) \left(\bar{\vec M}_i \cdot \vec M_j\right) + \text{c.c.}\right]$ coming from processes away from the patches.
 It can be used to determine the angle and relative phases when $\Delta_i=0$. However, the coupling terms when $\Delta_i\neq0$, which we consider here because of the degeneracy between CDW and SDW on the quadratic level, have much larger coefficients and also fix the relative angle as we show next.

\begin{figure}[t]
\center{\includegraphics[width=0.25\linewidth]{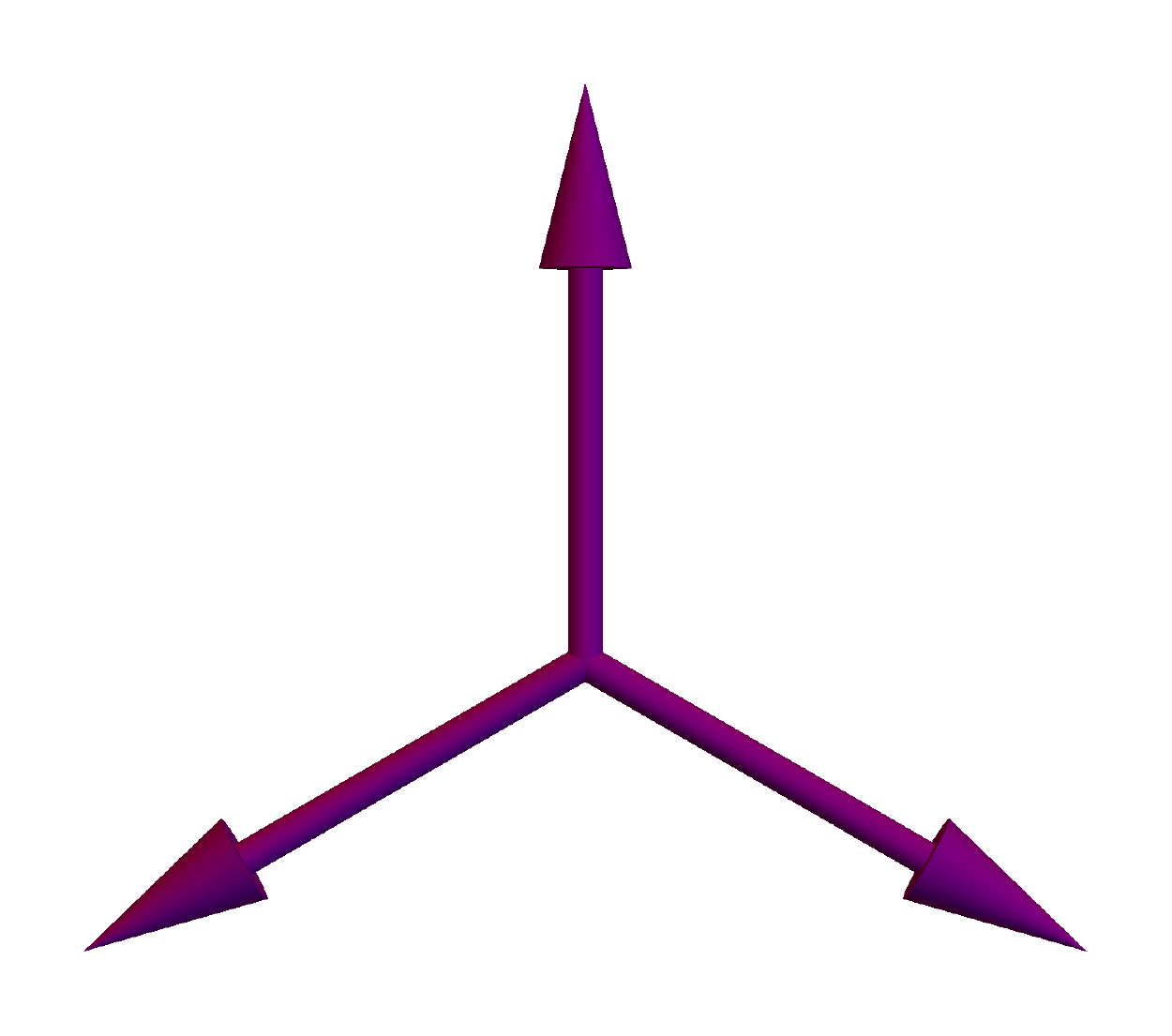} }
\centering{}\caption{The $120^o$ orientation of the SDW moments $\vec{M}_i$ in the ground state. The relative orientation of $\vec{M}_i$ is not specified in the SDW-only part of the free energy, but is determined by the coupling to CDW part.
}
\label{fig:mercedes}
\end{figure}

Motivated by our findings, we also parameterize $\Delta_i = \Delta e^{i \chi_i}$ in the coupling terms. Then we can rewrite the quartic part of the free energy in the form
\begin{align}
&\mathcal{F}_{cs} = 4Z_1 \Delta^2 M^2 [ \cos 2\gamma_1 + \cos 2\gamma_2 + \cos 2\gamma_3] \nonumber \\ &+ 8 Z_2 \Delta^2 M^2 [ \cos (\gamma_1 + \gamma_2) + \cos (\gamma_2 + \gamma_3) + \cos (\gamma_1 + \gamma_3) \nonumber \\ & + \cos (\gamma_1 - \gamma_2) \cos \theta_{12} + \cos (\gamma_2 - \gamma_3) \cos \theta_{23} + \cos (\gamma_1 - \gamma_3) \cos \theta_{13} ],
\end{align}
where $\gamma_i = \chi_i - \varphi_i$, and $\theta_{ij}$ is the angle between vectors $\vec{m}_i$ and $\vec{m}_j$. In 3D the sum of angles is constrained by
\begin{equation}
\theta_{12} + \theta_{23} + \theta_{13} \le 2\pi.
\end{equation}
 Because $Z_1 \gg Z_2 $, we can, to a good approximation, minimize
separately
 the parts of $\mathcal{F}_{cs}$ with $Z_1$ and with $Z_2$.  Minimizing the
 $Z_1$ part we find
\begin{equation}
\gamma_1 = \gamma_2 = \gamma_3 = \pm \frac{\pi}{2},
\end{equation}
 Minimizing then the $Z_2$ part, we find
 \begin{equation}
\theta_{12} = \theta_{23} = \theta_{13} = \frac{2\pi}{3}.
\end{equation}
  We also verified this result numerically.
The ground state structure is sketched in Fig.~\ref{fig:mercedes}.
In summary, we find that, in the ground state, the absolute values of the CDW and SDW fields are the same at all patch points, respectively. The relative phase between the complex CDW and SDW fields is $\pm \pi/2$, and the angle between the SDW moments is $2\pi/3$. The relative phases between fields at the different patch points is determined by processes away from the Fermi surface or higher-order terms in the free energy expansion.
The order parameter manifold is given by O(3)$\times$O(2)$\times$U(1)$\times$U(1)$\times$U(1)$\times Z_2$.
 The first $O(3) \times O(2)$ part is for the vectorial SDW component, U(1)$\times$U(1) is for the SDW and CDW components that break translational symmetry,  and U(1)$\times Z_2$  reflect the overall complex phase and two choices for the relative phase between the two orders.

\subsection{Ferromagnetic ground state}

For larger $\alpha_T$, the leading instability is towards $s$-wave spin Pomeranchuk order, i.e. intra-valley
ferromagnetism.
However, as we said, the relative orientation and the relative magnitude
of the order parameters ${\bf S}$ and ${\bf S}'$ in the two valleys are not determined at the quadratic level.
  To go beyond the quadratic level, we
   perform
 a Hubbard-Stratonovich transformation and integrate out the fermions.  We present the details in
    App.~\ref{app:HStechs} and here show the result.  We find
\begin{equation}
\mathcal{F}_{S}=\tilde\alpha
(1 - \lambda^s_{SPom})
 \left[\vec{S}\cdot \vec{S}  + \vec{S}'\cdot \vec{S}'\right]+2 \tilde Z_1
 \left[ (\vec{S} \cdot \vec{S})^2 + (\vec{S}' \cdot \vec{S}' )^2 \right]
  \label{eq:Ffm}
\end{equation}
 where
  $\tilde Z_1=T\sum_\omega\int d\vec k G_o^4(\vec k)$.
  The quartic term can be equally expressed as
  \beq
  \tilde Z_1
 \left[ (\vec{S} \cdot \vec{S}) + (\vec{S}' \cdot \vec{S}' ) \right]^2 +  \tilde Z_1
 \left[ (\vec{S} \cdot \vec{S}) - (\vec{S}' \cdot \vec{S}' ) \right]^2.
 \label{ex_1a}
  \eeq
 The first term sets the value of the square of the total order parameter, the second one
  sets the magnitudes of $\vec{S}$ and $|\vec{S}'|$ to be equal.  However, the relative orientation of
  $\vec{S}$ and $\vec{S}'$ is still undetermined. This degeneracy is the result of the decoupling between ferromagnetic order parameters from different valleys.

 We first check whether the degeneracy is lifted once we couple $\vec{S}$ and $\vec{S}'$ to fluctuating CDW and SDW  order parameters with momenta $\vec{Q}_s$, as  these order parameters couple fermions from different valleys.  The corresponding Landau functional is
\begin{align}\label{eq:Fcoupl}
\mathcal{F}_{coupl}=\mathcal{F}_{\Delta S} + \mathcal{F}_{M S} + \mathcal{F}_{\Delta M S}\,,
\end{align}
 (see App.~\ref{app:HStechs} for details). Here
\begin{align}
\mathcal{F}_{\Delta S}&=4Z_3\left(\vec S\cdot \vec S + \vec S' \cdot \vec S'  \right)\sum_i|\Delta_i|^2\ + 4 Z_1 (\vec{S} \cdot \vec{S'})\sum_i|\Delta_i|^2\\
\mathcal{F}_{M S}
&=4Z_3\left(\vec S\cdot \vec S + \vec S' \cdot \vec S'  \right)\sum_i \bar{\vec{M}}_i\cdot \vec M_i  - 4 Z_1 (\vec{S} \cdot \vec{S'})\sum_i \bar{\vec{M}}_i\cdot \vec M_i \nonumber \\ &+4 Z_1\sum_i \left[(\bar{\vec M}_i\cdot \vec S) (\vec M_i \cdot \vec S')+ (\bar{\vec M}_i\cdot \vec S') (\vec M_i \cdot \vec S) \right]\\
\mathcal{F}_{\Delta M S}&=
 -\frac{4}{3}K_3 (\vec S+ \vec S')\cdot\sum_i (\bar \Delta_i \vec M_i + \Delta_i \bar{\vec M}_i) +
 4 Z_1 (\vec S\times \vec S')\cdot\sum_i \left[i (\bar \Delta_i \vec M_i - \Delta_i \bar{\vec M}_i)\right]\,.
\label{eq:coupl3}
\end{align}
 where
  $Z_3 = T\sum_\omega \int d\vec k G_o(\vec k) G_p(\vec k +Q_s)^3$, and, we remind, $o$ and $p$ belong to different valleys, i.e., if $o$ is + then $p$ is $-$, and vice versa.
 The prefactor of the cubic term
   $K_3=T\sum_\omega \int d\vec k G_o^2(\vec k)G_p(\vec k+\vec Q_s)$ with $o \neq p$ vanishes within our approximation
   for $T\rightarrow 0$,
    but is finite if, e.g., we expand beyond quadratic level around Van Hove points.

\begin{figure}[t]
\includegraphics[width=0.6\linewidth]{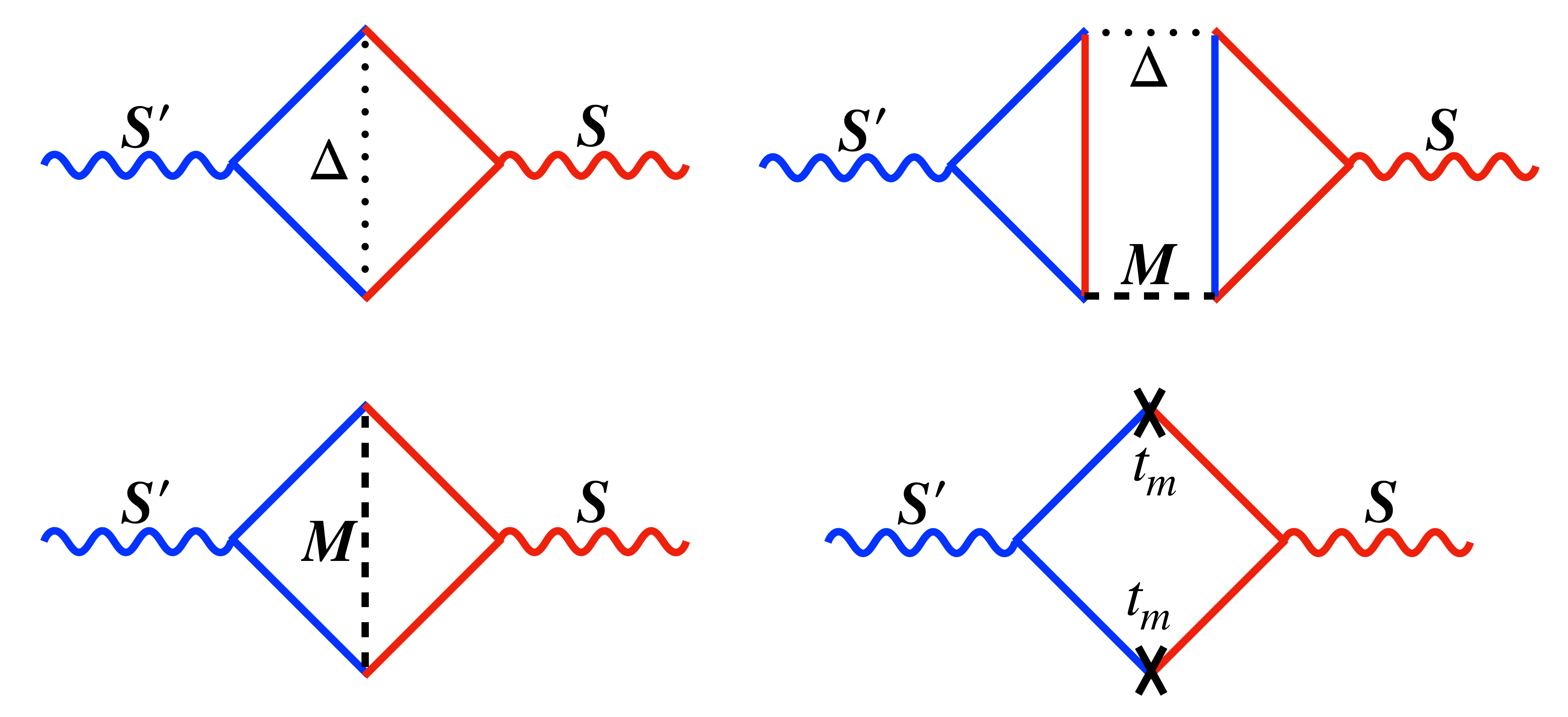}
\centering{}\caption{
The diagrams that contribute to
$ \bf S\cdot S'$ coupling between ferromagnetic order parameters in the two valleys.
 Solid red and blue lines denote fermionic propagators, wavy red and blue lines denote ferromagnetic order parameters $\bf S, S'$ in valleys $+$ and $-$.
  Dotted
  and dashed lines denote
 charge density wave and
 spin density wave fluctuations, respectively.
  We also show the diagram that
   gives  the coupling between $\bf S$ and $\bf S'$ due to
   inter-valley hopping $t_m$.}
\label{boxes}
\end{figure}

We obtain
 the leading
 contribution to the Landau functional $\mathcal{F}_{S}$ in (\ref{eq:Ffm})
 due to the coupling to the density wave fluctuations
by integrating out $\bar\Delta, \Delta$ and $\bar {\bf M}, {\bf M}$. The corresponding diagrams are presented in Fig.~\ref{boxes}.
 We find
\begin{align}
\mathcal{F}_{S,eff}=\mathcal{F}_{S}+
3 r \left[\vec{S}\cdot \vec{S}  + \vec{S}'\cdot \vec{S}'\right] + 3r'(\vec S\cdot \vec S')
\end{align}
 The prefactor for the first term is $r=8Z_3/\alpha-16K_3^2/(9\alpha^2)$.
  This term modifies the transition temperature, but does not couple order parameters from different valleys.  The  second term controls the relative orientation.
      However, the contributions to $r'$ from $\mathcal{F}_{\Delta S}$ and $\mathcal{F}_{M S}$  cancel each other:
      the one from CDW fluctuations  gives $4 Z_1/\alpha$ and
        taken alone
      would induce
      an antiferromagnetic coupling between valleys, but
      the one from SDW fluctuations  gives $-4 Z_1/\alpha$
      and
      would induce a
      ferromagnetic coupling.
        This cancellation is the consequence of the degeneracy between CDW and SDW fluctuations with
 momenta $\vec{Q}_s$.
 In principle, there is another contribution from the cubic term in $\mathcal{F}_{\Delta M S}$, which gives $r'=-32K_3^2/(9\alpha^2)$, but as we said before $K_3$
  is non-zero only if we go beyond  our patch model with quadratic expansion near the Van Hove points.
 The same holds if we couple ${\bf S}$ and ${\bf S}'$ to  CDW/SDW fluctuations with momenta ${\bf Q}_l$.   We also verified that the second order contribution to $\mathcal{F}_{S}$ from (\ref{eq:coupl3}) makes the prefactor for the last term in (\ref{ex_1a}) even more positive.

\begin{figure}[t]
\includegraphics[width=0.4\linewidth]{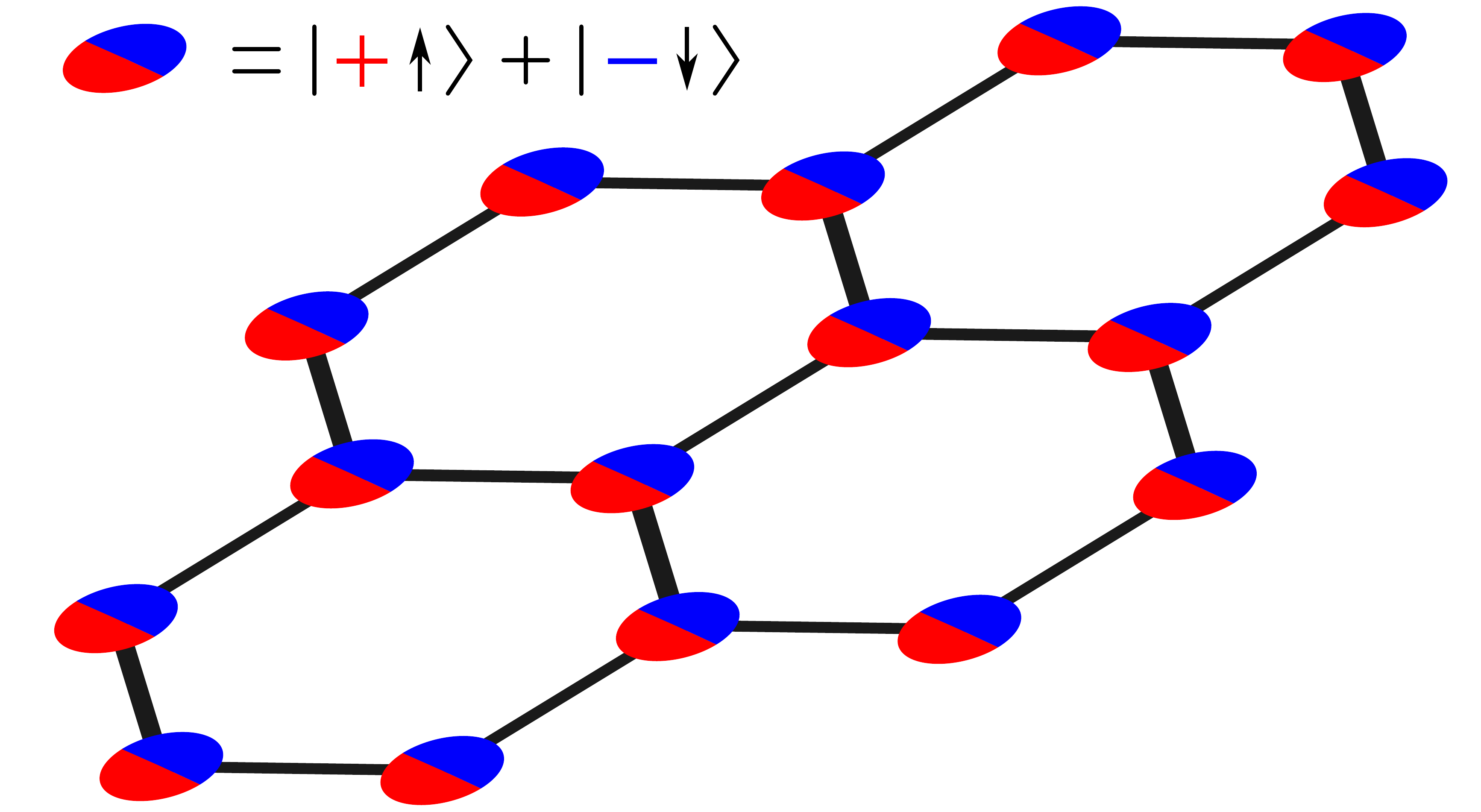}
\centering{}\caption{
The FM/AFM order parameter (intra-valley FM/inter-valley AFM)   in the real space.
  $+$ and $-$ denote the valleys on each superlattice site, and $|\uparrow\rangle$ and $|\downarrow\rangle$  represent two directions of the FM moments.}
\label{FMsk}
\end{figure}

A way to get a non-zero prefactor for the $\vec S\cdot \vec S'$ term
 within the patch model
is to include the hopping between valleys --  the one which  gives rise to valley mixing.  The hopping term relevant for this issue is $t_m c^+_{o,k} c_{p, k + Q_s} + h.c$.  Once we include this term, the charge
  contribution to $r'$ increases by $2Z_1 t_m^2$, 
  and $r'$ becomes non-zero and positive.
 As the consequence, $\vec{S}$ and $\vec{S'}$ order antiparallel to each other, and the resulting state is
 an intra-valley FM and inter-valley AFM (FM/AFM state). Spins of different valleys point in opposite directions on every site of the superlattice, as sketched in Fig. \ref{FMsk}. Such a state has no net magnetization.

The FM/AFM state is identical to the one found in Refs.~\onlinecite{Kang2018strong}
   within a strong-coupling analysis. Furthermore, the mechanism that lifts the degeneracy between different valleys in our itinerant approach is similar to the one in the strong-coupling scenario. In both cases, valley-mixing terms favor antiferromagnetic ordering of magnetic moments from different  valleys.
   The fact that both weak- and strong-coupling approaches give the same result suggests that FM/AFM order is quite robust and likely survives at all couplings (see Ref.~\cite{Vafek2010weakstrong} for a similar situation in bilayer graphene).

\subsection{d-wave Pomeranchuk order}
\label{sec:dPoms}
In our case  the $d$-wave spin and charge Pomeranchuk channels are also attractive, see Fig.~\ref{lambda_alpha}.
 A $d$-wave Pomeranchuk order can additionally break lattice rotational symmetry, and we explore the possibility that the ordered state is a nematic (or that there are strong nematic fluctuations, if the eigenvalue is below the threshold for the instability).

 We remind that the eigenvalues in the $d$-wave charge and spin Pomeranchuk channels are degenerate, and there is also valley degeneracy.  Furthermore, each order parameter has two components as it belongs to the two-dimensional representation $E$
 of the $D_3$ symmetry group of the Hamiltonian.
Accordingly, we introduce two-component scalar charge $d-$wave Pomeranchuk order parameters $(\chi_{d1}, \chi_{d2})$ and $(\chi'_{d1}, \chi'_{d2})$ and two two-component vector spin $d-$wave Pomeranchuk order parameters $(\vec\phi_{d1}, \vec\phi_{d2})$ and $(\vec\phi'_{d1}, \vec\phi'_{d2})$
 (see Eqs.~\eqref{eq:opCPom} and \eqref{eq:opSPom}).

 Performing the Hubbard-Stratonovich transformation and integrating out fermions, we obtain the free energy in terms of
   $\chi$ and $\vec\phi$:
\beq
\mathcal{F}_d=\mathcal{F}^{(2)}_d+\mathcal{F}^c_d+\mathcal{F}^s_d+\mathcal{F}^{cs}_d
\eeq
with the quadratic part given by Eq.~\eqref{eq:quadratic_dPOM}, and with
\begin{align}
\mathcal{F}^{c}_d&=-\frac{\sqrt{2}}{3\sqrt{3}}\tilde K_3 (\chi_{d1}^3-3\chi_{d1}\chi_{d2}^2) + \frac{1}{4}\tilde Z_1(\chi_{d1}^2+\chi_{d2}^2)^2 + \{\chi_i\leftrightarrow \chi'_i\}\\
\mathcal{F}^{s}_d&=\frac{1}{4}\tilde Z_1 \left[ (\vec\phi_{d1}^2+\vec\phi_{d2}^2)^2-\frac{2}{3}\vec\phi_{d1}^2\vec\phi_{d2}^2+\frac{2}{3}(\vec\phi_{d1}\cdot\vec\phi_{d2})^2 \right] +\{\vec\phi_i\leftrightarrow\vec\phi'_i\}\\
\mathcal{F}^{cs}_d&=-\sqrt{\frac{2}{3}}\tilde K_3 \left[ (\vec \phi_{d1}^2-\vec\phi_{d2}^2)\chi_{d1} -2 (\vec \phi_{d1}\cdot\vec \phi_{d2}) \chi_{d2}\right]\nonumber\\
&+\frac{1}{4}\tilde Z_1 \left[ (\vec\phi_{d1}^2+\vec\phi_{d2}^2)(\chi_{d1}^2+\chi_{d2}^2) -\frac{2}{3}\vec\phi_{d1}^2\chi_{d2}^2 -\frac{2}{3}\vec\phi_{d2}^2\chi_{d1}^2 +\frac{4}{3}(\vec\phi_{d1}\cdot\vec\phi_{d2})\chi_{d1}\chi_{d2}\right]+ \{\chi_i\leftrightarrow \chi'_i, \vec\phi_i\leftrightarrow\vec\phi'_i\}
\label{eq:FCSPom}
\end{align}
where $\tilde K_3=T\sum_\omega\int d\vec k G_o^3(k)$ and $\tilde Z_1$ are defined below Eq.~\eqref{eq:Ffm}.
We see that  the free energy contains cubic terms with the form $\chi^3$ and $\vec\phi^2 \chi$. The cubic terms can be re-expressed as $\mathcal F_d^{(3)}=-\sqrt{2/3}\tilde K_3 \left[ (\chi_{d1}+i\chi_{d2})^3/6+(\chi_{d1}+i\chi_{d2})(\vec\phi_{d1}+i\vec\phi_{d2})^2/2 + \text{c.c.}\right]+ \{\chi_i\leftrightarrow \chi'_i, \vec\phi_i\leftrightarrow\vec\phi'_i\}$, which makes the symmetry under threefold rotations more apparent.
The presence of the cubic terms indicates that the transition to the d-wave Pomeranchuk order is first order.
The contributions to $\mathcal F_d^{(3)}$ from different valleys are decoupled, which is again a consequence of the absence of valley mixing.

We first analyze spin and charge parts of the free energy, $\mathcal{F}^c_d$ and $\mathcal{F}^s_d$, separately, neglecting the coupling term $\mathcal{F}^{cs}_d$.
A straightforward analysis shows that the free energy for the $d$-wave charge Pomeranchuk order $\mathcal{F}^c_d$ is minimized by one of the three configurations
\begin{align}
(\chi_{d1},\chi_{d2})=\chi(1,0) \qquad (\chi_{d1},\chi_{d2})=\frac{\chi}{2}(-1,\sqrt{3})  \qquad (\chi_{d1},\chi_{d2})=\frac{\chi}{2}(-1,-\sqrt{3})
\label{l1}
\end{align}
and analogously in the other valley sector
\begin{align}
(\chi'_{d1},\chi'_{d2})=\chi'(1,0) \qquad (\chi'_{d1},\chi'_{d2})=\frac{\chi'}{2}(-1,\sqrt{3})  \qquad (\chi'_{d1},\chi'_{d2})=\frac{\chi'}{2}(-1,-\sqrt{3})
\label{l2}
\end{align}
The system spontaneously chooses one of these minima, i.e. a certain charge distribution in the Van Hove patches. This breaks the threefold rotation symmetry and leads to a nematic order.
In real space, the $d$-wave form factor leads to a modulation of hopping amplitudes.  For each choice of one of the states from (\ref{l1}) and (\ref{l2}), the threefold rotation symmetry gets broken. Without any coupling between the two valley sectors, any combination of the minima in the two sectors is equivalent. Four of the nine possible combinations also spontaneously break the symmetry between the valleys.
Valley-mixing terms have to be introduced to determine which configuration minimizes the free energy. This can be done either by adding extra terms to the single-particle Hamiltonian, or as in the previous section, by analyzing the effects of the coupling to fluctuations of order parameters from different channels.

The free energy for the $d$-wave spin Pomeranchuk order $\mathcal{F}^s_d$ does not contain cubic terms.
Within each valley, it is minimized by setting
 $\vec\phi_{d1}^2=\vec\phi_{d2}^2$, $(\vec\phi'_{d1})^2=(\vec\phi'_{d2})^2$ and  $\vec\phi_{d_1}\cdot\vec\phi_{d2}=0$, $\vec\phi'_{d_1}\cdot\vec\phi'_{d2}=0$.
 Such an order has recently been studied in Ref.~\onlinecite{2020arXiv200614729C}. The total spin order parameter with this configuration winds twice around the unit circle. It breaks the spin SU(2) symmetry and introduces a Zeeman-like splitting in the energy dispersion. However, because of the $d$-wave form factor, there is no net magnetization. In real space, the $d$-wave form factor again modulates the hopping amplitudes, but now the hopping modulation becomes spin-dependent. The relative orientation between order parameters in different valley sectors remains undetermined at this level due to the absence of valley mixing and is again set by either adding valley-mixing terms to the single-particle Hamiltonian, or by analyzing the effects of the coupling to fluctuations of order parameters from different channels.

We now include into consideration the term $\mathcal{F}^{cs}_d$, which couples $d$-wave charge and spin Pomeranchuk orders.
It introduces cubic terms that are linear in the charge order parameters and quadratic in the spin order parameters, cf. Eq.~(\ref{eq:FCSPom}). We assume that the magnitudes of the order parameters are small. Then the cubic terms are more important than the quartic terms. In this case, the nematic charge order forces the spin order to also become a nematic.  Indeed,
let's  focus on a particular valley sector and
choose the state $\chi(1,0)$ in the charge sector.
 Substituting the corresponding $\chi$ into the coupling term, we obtain
\beq
\mathcal F_d^{cs,(3)}=-\sqrt{\frac{2}{3}}\tilde K_3 \chi (\vec\phi_{d1}^2-\vec\phi_{d2}^2).
\eeq
This free energy
 favors
\beq
(|\vec\phi_{d1}|, |\vec \phi_{d2}|)=\phi(1,0).
\eeq
For the other two nematic charge orders  $(\chi_{d_1},\chi_{d2})=\chi/2(-1,\pm\sqrt{3})$, we obtain
\beq
\mathcal F_d^{cs,(3)}=\sqrt{\frac{2}{3}}\tilde K_3 \frac{\chi}{2}\left[ (\vec\phi_{d1}^2-\vec\phi_{d2}^2)\pm2(\vec\phi_{d1}\cdot\vec\phi_{d2})\right]
\eeq
The sign of the last term determines if $\vec\phi_{d1}$ and $\vec\phi_{d2}$ align parallel or antiparallel. In both cases, the magnitudes of $\vec \phi_{d1}$ and $\vec\phi_{d2}$ become
\beq
(|\vec\phi_{d1}|, |\vec \phi_{d2}|)=\frac{\phi}{2}(1,\sqrt{3})\,.
\eeq
We see therefore that, at least when the magnitudes of the  order parameters are small, the nematic order in the charge sector induces nematic order in the spin sector.
Whether the nematic order in the spin channel persists at larger $\vec\phi$ depends on the interplay between cubic and quartic terms in the free energy for the spin order parameter. Also, as before, it depends on the coupling between the two valley sectors, if the nematic combination additionally breaks the valley symmetry or not.

\section{Density-wave and Pomeranchuk orders in the 12-patch model}


\begin{figure}[h]
\includegraphics[width=0.5\linewidth]{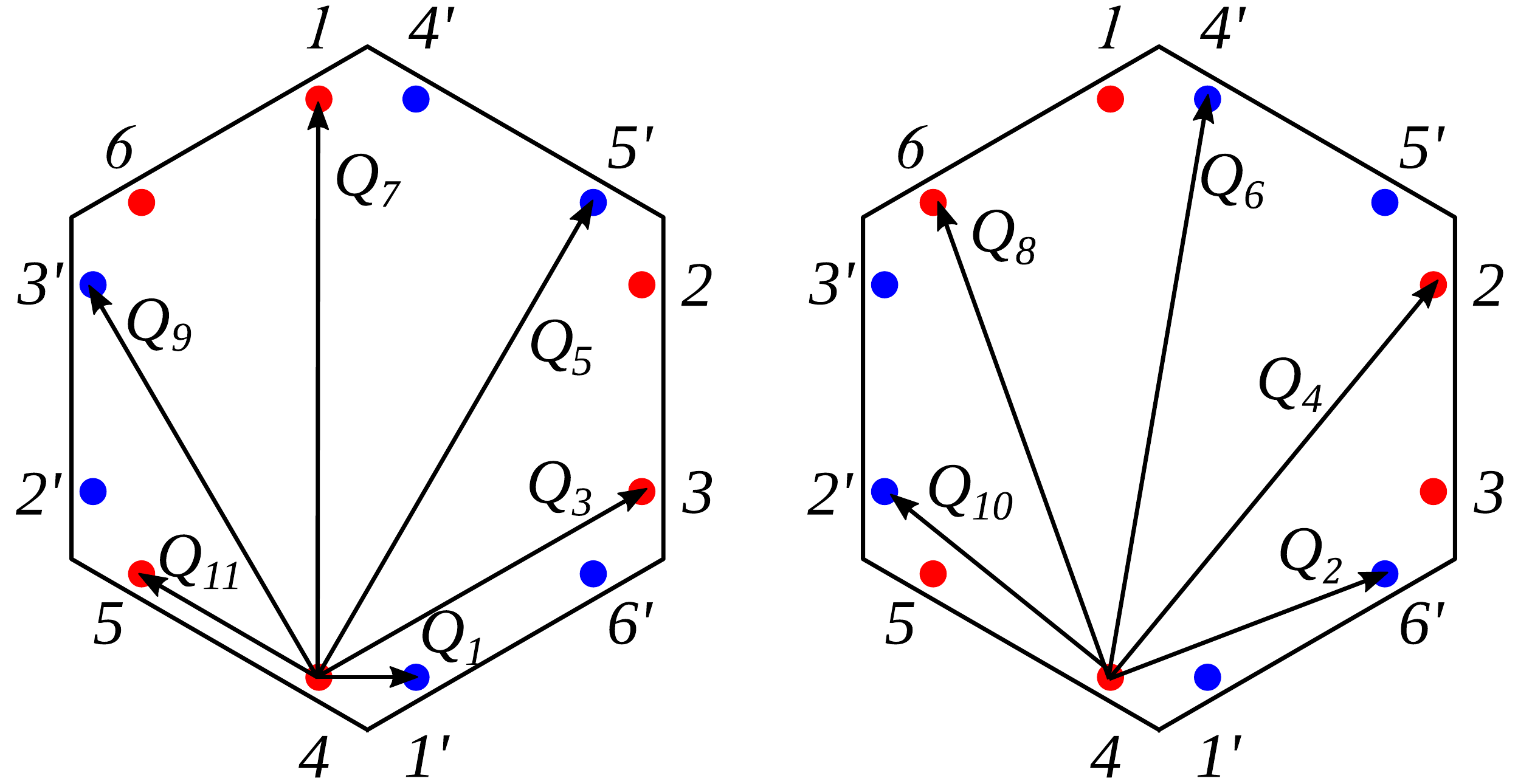}
\caption{ Possible momentum transfers between low-energy fermions in the twelve-patch model. Blue and red dots mark  Van Hove points for fermions from one or the other valley. In the twelve-patch model, there are eleven different types of momentum transfers, ${\bf Q}_1 - {\bf Q}_{11}$, however $Q_4$ and $Q_8$, and $Q_2$ and $Q_{10}$ are equivalent due to a rotation symmetry.
 }
\label{mod_sk12p}
\end{figure}


\label{sec:12patch}

We now proceed with the analysis of particle-hole orders in the twelve-patch model. We follow the same strategy as in previous sections, i.e., we introduce all possible vertices involving one incoming and one outgoing fermion and consider their renormalizations within the ladder approximation.

\subsection{The polarization bubbles}
\label{sec:12polar}

Like in
the six patch model, a half of the patches is formed by fermions with one valley index, and a half by fermions with the other valley index. There are eleven possible,
non-zero
 momentum transfers. However  $\vec{Q}_4$ and $\vec{Q}_8$, and $\vec{Q}_2$ and $\vec{Q}_{10}$ are related by $C_3$ symmetry ( $|\vec Q_4|=|\vec Q_8|$ and $|\vec Q_2|=|\vec Q_{10}|$),
therefore the actual number of different momentum transfers is nine.

 We introduce inter-valley $\Pi_{+-} (Q_j)=-\int G_{+} (k) {G}_{-} (k+Q_j)>0$ and intra-valley $\Pi_{++} (Q_j)=-\int G_{+} (k) G_{+} (k+Q_j)>0$.
   The symmetry constraints are the same as before: $\Pi_{++} (Q_j)=\Pi_{--} (Q_j)$ and $\Pi_{+-} (Q_j)=\Pi_{-+} (Q_j)$.
     In distinction to the six-patch model, there are now more than one intra-valley  $\Pi_{++} (Q_j)$.
      All polarization bubbles are logarithmically divergent at the Van Hove doping. We
       choose $\Pi_{++} (0)$ as the basic one and  express all polarization bubbles in units of $\Pi_{++} (0)$.
     We present the results in Table \ref{polar_table_12p}.

\begin{table}[h!]
\centering
 \begin{tabular}{||c | c | c||}
 \hline
 Polarization operator $\Pi_{op} (\vec{Q})$ & Green's functions & $\Pi_{op} (\vec{Q}) / \Pi_{++} (0)$  \\  [1ex]
 \hline \hline
 $\Pi_{++} (0)$ & $-\int G_+(k) G_+(k)$ & 1 \\
 \hline
 $\Pi_{+-} (\vec{Q}_1)$ & $-\int G_+(k) G_-(k+Q_1)$ & 1.32  \\
 \hline
 $\Pi_{+-} (\vec{Q}_2) = \Pi_{+-} (\vec{Q}_{10}) $ & $-\int G_+(k) G_-(k+Q_2)$ &  1.07 \\
 \hline
 $\Pi_{++} (\vec{Q}_3)$ & $-\int G_+(k) G_+(k+Q_3)$ & 0.72  \\
 \hline
  $\Pi_{++} (\vec{Q}_4) = \Pi_{++} (\vec{Q}_8) $ & $-\int G_+(k) G_+(k+Q_4)$ & 1.1 \\
 \hline
  $\Pi_{+-} (\vec{Q}_5)$ & $-\int G_+(k) G_-(k+Q_5)$ & 1.36 \\
 \hline
  $\Pi_{+-} (\vec{Q}_6)$ & $-\int G_+(k) G_-(k+Q_6)$ & 0.75 \\
 \hline
  $\Pi_{++} (\vec{Q}_7)$ & $-\int G_+(k) G_+(k+Q_7)$ & 0.81  \\
 \hline
  $\Pi_{+-} (\vec{Q}_9)$ & $-\int G_+(k) G_-(k+Q_9)$ & 1.09 \\
 \hline
  $\Pi_{++} (\vec{Q}_{11})$ & $-\int G_+(k) G_+(k+Q_{11})$ & 0.73 \\
 \hline
\end{tabular}
\centering{}\caption{Intra-valley $\Pi_{++}$ and inter-valley $\Pi_{+-}$ polarization bubbles, in units of
 $\Pi_{++} (0)$, at $T=0$. Like before, we moved the chemical potential slightly away from Van Hove doping by
 $\delta \mu \sim 0.001$ to  regularize logarithmic divergencies.
$G_\pm(k)=1/(i\omega-E_\pm(k))$ are the Green's functions of fermions from different valleys (bands).}
\label{polar_table_12p}
\end{table}

\subsection{The dressed vertices}
\label{sec:12RPA}


\begin{figure}[h]
\center{\includegraphics[width=0.99\linewidth]{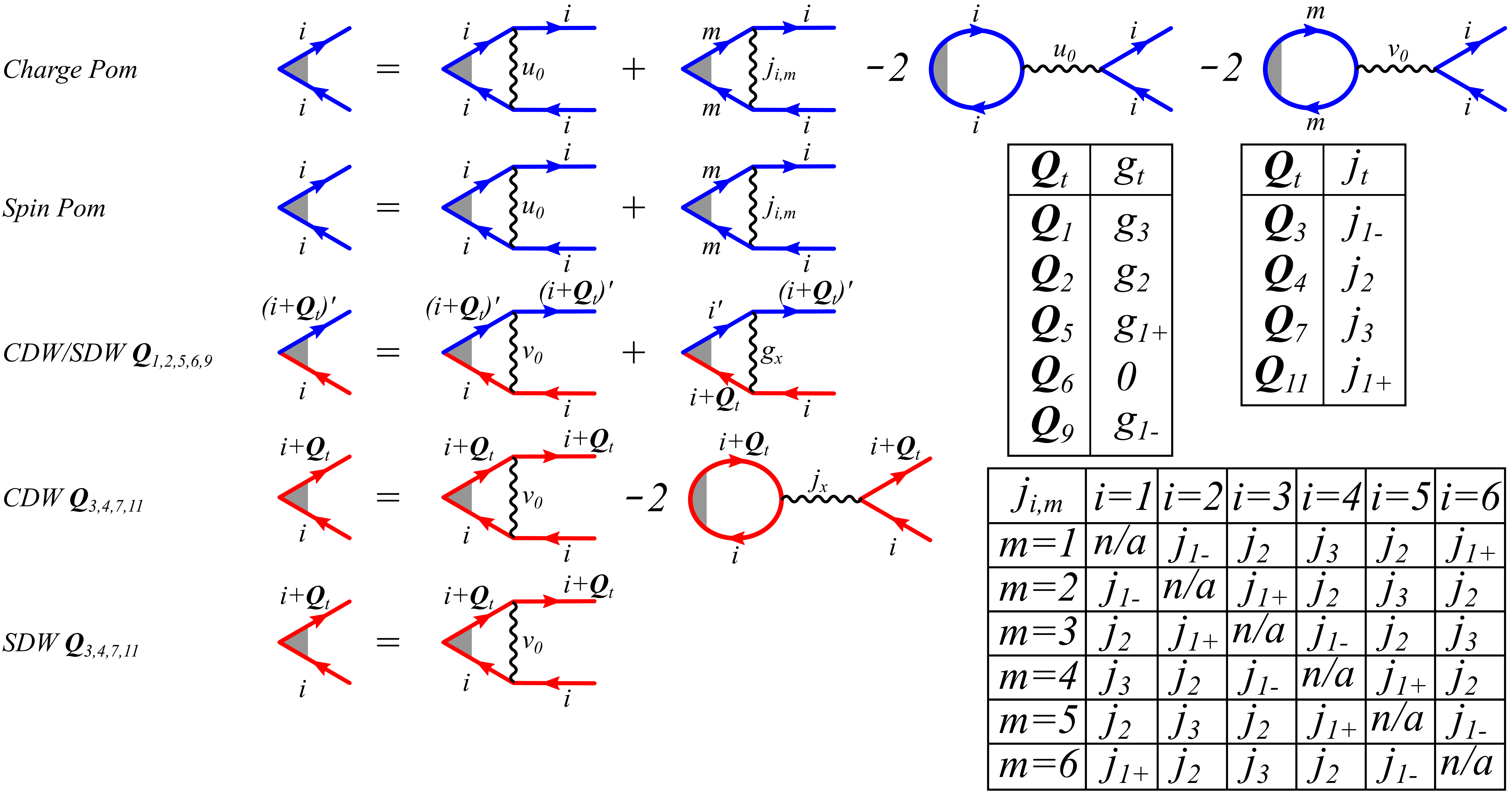} }
\centering{}\caption{Diagrammatic representation of a system of coupled gap equations. Gray triangle is a fully renormalized vertex, red and blue lines are Green's functions of  electrons from the two bands. Summation over $m \neq i$ is implied. When diagrammatic equation involves only one color of fermions, it is implied that there is another identical equation for the other color.  Two upper tables show the values of couplings for orders with the corresponding momentum transfer. The lower table show values of couplings for Pomeranchuk orders.
}
\label{gapeq_pic_12p}
\end{figure}


A straightforward analysis shows that the number of order parameters (fermionic bilinears)  is 20: nine different CDW orders and nine different SDW orders with various momenta, and spin and charge Pomeranchuk  orders  with $\vec{Q}=0$.
We list the spin order parameters in Table \ref{order_table_12p}. The charge order parameters are obtained by substituting $\vec{\sigma}$ by Kronecker delta $\delta_{ss'}$. We introduce trial vertices with the structure of these order parameters and  write down matrix equations for the dressed vertices
that include corrections from interactions.
 We diagonalize these equations, obtain dimensionless couplings, and identify
 the channel with the largest attractive coupling. The equations for the dressed vertices  are shown schematically  in Fig. \ref{gapeq_pic_12p}.
\begin{table}[h]
\centering
\caption{List of order parameters  in the spin channel in the twelve-patch model. Here $\vec{\sigma}$ is the vector of Pauli matrices.  All spin order parameters are vectors. Order parameters in the charge channel are obtained by substituting $\vec{\sigma}$ by Kronecker delta $\delta_{ss'}$. }
\begin{adjustbox}{width=\columnwidth,center}
 \begin{tabular}{||c | c | c | c| c|c||}
 \hline
 Order & Vertex  & Patch order parameters & Fermionic bilinear & Number of fields & Real or complex \\  [1ex]
 \hline \hline
 Spin $\vec{Q}=0$ (Pom) & $\Gamma_s(0)$& $\Delta^s_i (0) $ and $\Delta^s_{i'} (0) $ & $\av{f^{\dagger}_{is} \vec{\sigma}_{ss'} f_{is'}}$ and $\av{f^{\dagger}_{i's} \vec{\sigma}_{ss'} f_{i's'}}$ & 12 & Real \\
 \hline
 Spin $\vec{Q}_1$ & $\Gamma_s(\vec Q_1)$ & $\Delta^s_i (\vec{Q}_1) $ and $\Delta^s_{i'} (\vec{Q}_1) $ & $\av{f^{\dagger}_{i's} \vec{\sigma}_{ss'} f_{(i+3)s'}}$ and $\av{f^{\dagger}_{(i+3)'s} \vec{\sigma}_{ss'} f_{is'}}$,  $i=1..3$  & 6 & Complex\\
 \hline
 Spin $\vec{Q}_2$ &$\Gamma_s(\vec Q_2)$ & $\Delta^s_i (\vec{Q}_2) $ and $\Delta^s_{i'} (\vec{Q}_2) $ & $\av{f^{\dagger}_{(i+2)'s} \vec{\sigma}_{ss'} f_{is'}}$ and $\av{f^{\dagger}_{i's} \vec{\sigma}_{ss'} f_{(i+2)s'}}$ & 12 & Complex\\
 \hline
  Spin $\vec{Q}_3$ &$\Gamma_s(\vec Q_3)$ & $\Delta^s_i (\vec{Q}_3) $ and $\Delta^s_{i'} (\vec{Q}_3) $ & $\av{f^{\dagger}_{is} \vec{\sigma}_{ss'} f_{(i+1)s'}}$ and $\av{f^{\dagger}_{i's} \vec{\sigma}_{ss'} f_{(i+1)'s'}}$,
  $i=odd$ & 6 & Complex\\
 \hline
  Spin $\vec{Q}_4$ &$\Gamma_s(\vec Q_4)$ & $\Delta^s_i (\vec{Q}_4) $ and $\Delta^s_{i'} (\vec{Q}_4) $ & $\av{f^{\dagger}_{is} \vec{\sigma}_{ss'} f_{(i+2)s'}}$ and $\av{f^{\dagger}_{i's} \vec{\sigma}_{ss'} f_{(i+2)'s'}}$  & 12 & Complex\\
 \hline
 Spin $\vec{Q}_5$ &$\Gamma_s(\vec Q_5)$ & $\Delta^s_i (\vec{Q}_5) $ and $\Delta^s_{i'} (\vec{Q}_5) $ & $\av{f^{\dagger}_{(i+1)'s} \vec{\sigma}_{ss'} f_{is'}}$ and $\av{f^{\dagger}_{i's} \vec{\sigma}_{ss'} f_{(i+1)s'}}$,  $i=even$  & 6 & Complex\\
 \hline
 Spin $\vec{Q}_6$ &$\Gamma_s(\vec Q_6)$ & $\Delta^s_i (\vec{Q}_6)
 ={\Delta^s_{i'}} (\vec{Q}_6)^\dagger
 $  & $\av{f^{\dagger}_{i's} \vec{\sigma}_{ss'} f_{is'}}$ & 6 & Complex\\
 \hline
  Spin $\vec{Q}_7$ &$\Gamma_s(\vec Q_7)$ & $\Delta^s_i (\vec{Q}_7) $ and $\Delta^s_{i'} (\vec{Q}_7) $ & $\av{f^{\dagger}_{is} \vec{\sigma}_{ss'} f_{(i+3)s'}}$ and $\av{f^{\dagger}_{(i+3)'s} \vec{\sigma}_{ss'} f_{i's'}}$,  $i=1..3$  & 6 & Complex\\
 \hline
  Spin $\vec{Q}_9$ &$\Gamma_s(\vec Q_9)$ & $\Delta^s_i (\vec{Q}_9) $ and $\Delta^s_{i'} (\vec{Q}_9) $ & $\av{f^{\dagger}_{(i-1)'s} \vec{\sigma}_{ss'} f_{is'}}$ and $\av{f^{\dagger}_{i's} \vec{\sigma}_{ss'} f_{(i-1)s'}}$,  $i=even$  & 6 & Complex\\
 \hline
  Spin $\vec{Q}_{11}$ &$\Gamma_s(\vec Q_{11})$ & $\Delta^s_i (\vec{Q}_{11}) $ and $\Delta^s_{i'} (\vec{Q}_{11}) $ & $\av{f^{\dagger}_{(i+1)s} \vec{\sigma}_{ss'} f_{is'}}$ and $\av{f^{\dagger}_{i's} \vec{\sigma}_{ss'} f_{(i+1)'s'}}$,  $i=even$  & 6 & Complex\\
 \hline
\end{tabular}
\end{adjustbox}
\label{order_table_12p}
\end{table}

For Pomeranchuk channels, $\vec{Q}=0$ orders for different valleys are decoupled.
The ladder series for the dressed  Pomeranchuk vertices  yield (see Fig.~\ref{gapeq_pic_12p})
\begin{align}
\Gamma_{CPom}(0)&=\Gamma_{CPom}^0(0)+\Pi_{++}(0)\Lambda_{CPom,0}\Gamma_{CPom}(0)\\
\Gamma_{SPom}(0)&=\Gamma_{SPom}^0(0)+\Pi_{++}(0)\Lambda_{SPom,0}\Gamma_{SPom}(0)
\end{align}
where
\begin{equation}
\Lambda_{SPom,0}=\mathbbm{1}_{2 \times 2}\otimes
\begin{pmatrix}
u & g_{1-} & g_2 & g_3 & g_2 & g_{1+} \\
g_{1-} & u & g_{1+} & g_2 & g_3 & g_2 \\
g_2 & g_{1+} & u & g_{1-} & g_2 & g_3 \\
g_3 & g_2 & g_{1-}  & u & g_{1+} & g_2 \\
g_2 & g_3 & g_2 & g_{1+} & u & g_{1-} \\
g_{1+} & g_2 & g_3 & g_{2} & g_{1-} & u
\end{pmatrix};
\; \; \;
\Lambda_{CPom,0}=\mathbbm{1}_{2 \times 2}\otimes  \Lambda_{SPom,0} -  \mathbbm{1}_{2 \times 2}\otimes 2u
\begin{pmatrix}
1 & 1 & 1 & 1 & 1 & 1 \\
1 & 1 & 1 & 1 & 1 & 1 \\
1 & 1 & 1 & 1 & 1 & 1 \\
1 & 1 & 1  & 1 & 1 & 1 \\
1 & 1 & 1 & 1 & 1 & 1 \\
1 & 1 & 1 & 1 & 1 & 1
\end{pmatrix}.
\label{Pom_matr}
\end{equation}
Here $\mathbbm{1}_{2 \times 2}$ is a $2 \times 2$ unit matrix, acting in the valley space.
The five couplings $u$, $ g_{1+}$, $g_2, g_3$, and $g_{1-}$ are presented in Eq. (\ref{coupl_2}).

Density wave vertices can be either intra-valley (connecting patches, where low-energy excitations are made of fermions from the same valley) or inter-valley (connecting patches where low-energy fermions are from different valleys). Intra-valley density wave orders involve momentum transfers $\vec{Q}_3,\vec{Q}_4,\vec{Q}_7,\vec{Q}_{11}$, and inter-valley density wave orders are for momenta  $\vec{Q}_1,\vec{Q}_2,\vec{Q}_5,\vec{Q}_6,\vec{Q}_9$. The dressed vertices for intra-valley CDW and SDW are of the generic form
\begin{align}
\Gamma_{CDW}(\vec Q)=\Gamma_{CDW}^0(\vec Q)  +\Pi_{++}(\vec Q)\Lambda_{CDW \, Q}\Gamma_{CDW}(\vec Q) \\
\Gamma_{SDW}(\vec Q)=\Gamma_{SDW}^0(\vec Q)  +\Pi_{++}(\vec Q)\Lambda_{{SDW}\, Q}\Gamma_{SDW}(\vec Q),
\end{align}
where the  matrices $\Lambda_{CDW \, Q}$ and $\Lambda_{SDW \, Q}$ are
 block-diagonal due to the absence of valley mixing.  For the
 intra-valley
 CDW channels, we have
\begin{align}
\label{eq:CDWQ3}
\Lambda_{CDW}(\vec Q_3)&=(u-2g_{1-})\mathbbm{1}_{6 \times 6}\\
\Lambda_{CDW}(\vec Q_4)&=(u-2g_2)\mathbbm{1}_{12 \times 12}\\
\Lambda_{CDW}(\vec Q_7)&=(u-2g_3)\mathbbm{1}_{6 \times 6}\\
\Lambda_{CDW}(\vec Q_{11})&=(u-2g_{1+})\mathbbm{1}_{6 \times 6}\,,
\label{eq:CDWQ11}
\end{align}
and for
 intra-valley
 SDW channels  the matrices are
\begin{equation}
\label{eq:intraSDW}
\Lambda_{SDW}(\vec Q_3)=\Lambda_{SDW}(\vec Q_7)=\Lambda_{SDW}(\vec Q_{11})=u\mathbbm{1}_{6 \times 6}; \; \Lambda_{SDW}(\vec Q_4)=u\mathbbm{1}_{12 \times 12},
\end{equation}
where $\mathbbm{1}_{i \times i}$ is a $i \times i$ unit matrix, reflecting
the diagonal forms of the matrix equations.
For inter-valley vertices with momenta $\vec{Q}_1,\vec{Q}_2,\vec{Q}_5,\vec{Q}_6,\vec{Q}_9$ the ladder series do not distinguish between SDW and CDW channels, because the diagrams that would break the equivalence between SDW and CDW are absent in the absence of valley mixing (see Fig.~\ref{gapeq_pic_12p}). As a result,
$\Lambda_{SDW }(\vec Q_i)=\Lambda_{CDW }(\vec Q_i)$.
We find for the different $\vec Q_i$
\begin{equation}
\begin{gathered}
\Lambda_{SDW }(\vec Q_1)=\Lambda_{CDW }(\vec Q_1)=\begin{pmatrix}
u & 0 & 0 & g_3 &0 &0 \\
0 & u & 0  &0 & g_3 &0\\
0 & 0 & u & 0 &0 & g_3 \\
g_3 & 0& 0 &u & 0&0 \\
0& g_3 & 0 &0 &u & 0 \\
0& 0& g_3 & 0& 0 & u
\end{pmatrix}; \; \; \;
\Lambda_{SDW }(\vec Q_2)=\Lambda_{CDW }(\vec Q_2) = u  \mathbbm{1}_{12 \times 12} + \sigma_1 \otimes g_2 \mathbbm{1}_{6 \times 6}  \\
\Lambda_{SDW }(\vec Q_5)=\Lambda_{CDW }(\vec Q_5)=\begin{pmatrix}
u & 0 & 0 & g_{1+} &0 &0 \\
0 & u & 0  &0 & g_{1+} &0\\
0 & 0 & u & 0 &0 & g_{1+} \\
g_{1+} & 0& 0 &u & 0&0 \\
0& g_{1+} & 0 &0 &u & 0 \\
0& 0& g_{1+} & 0& 0 & u
\end{pmatrix}; \; \; \;
\Lambda_{SDW }(\vec Q_9)=\Lambda_{CDW }(\vec Q_9)=\begin{pmatrix}
u & 0 & 0 & g_{1-} &0 &0 \\
0 & u & 0  &0 & g_{1-} &0\\
0 & 0 & u & 0 &0 & g_{1-} \\
g_{1-} & 0& 0 &u & 0&0 \\
0& g_{1-} & 0 &0 &u & 0 \\
0& 0& g_{1-} & 0& 0 & u
\end{pmatrix} \\
\Lambda_{SDW}(\vec Q_6)=\Lambda_{CDW }(\vec Q_6)=u \mathbbm{1}\,.
\end{gathered}
\label{12p_inter}
\end{equation}
where $\sigma_1$ acts in the
space.
 The matrices $\Lambda$ are either block-diagonal, or can be made block-diagonal by permutations of rows and columns.

\subsection{The eigenvalues}
\label{sec:12EIGEN}

\begin{figure}[h]
\center{\includegraphics[width=0.99\linewidth]{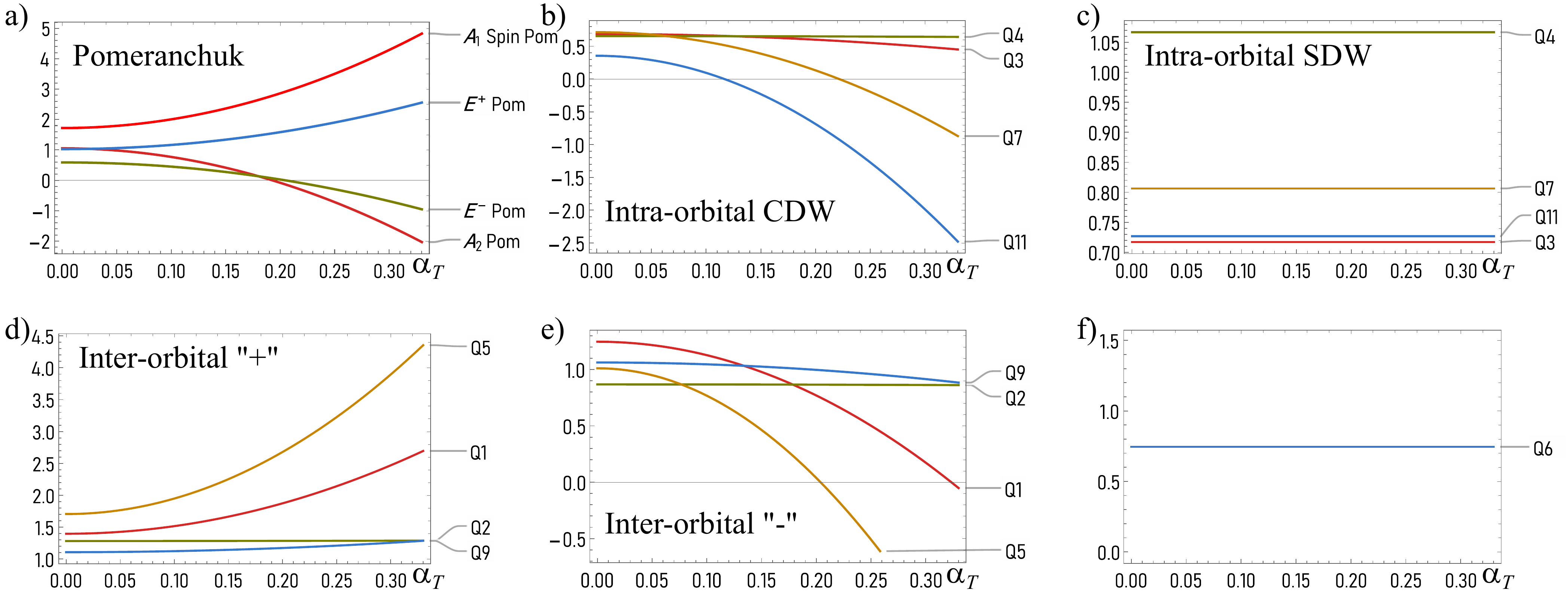} }
\centering{}\caption{Eigenvalues of the 12-patch model as  functions of $\alpha_T$.  A positive value of an eigenvalue means an attraction in the corresponding channel. a) Spin and  charge Pomeranchuk channels. The $A_1$ charge channel is omitted because it is strongly repulsive.  b) CDW channels with intra-valley polarization bubble.  c) SDW channels with intra-valley polarization bubble. d) Symmetric ( $+$) CDW/SDW channels,  e) Antisymmetric ( $-$) CDW/SDW channels with inter-valley polarization bubble.  f)
CDW/SDW channels with momentum transfer $\vec{Q}_6$. All eigenvalues are normalized to $\Pi_{++} (0)$.
 The numbers are in units of $V_0$.}
\label{lambda_alpha_12p}
\end{figure}

  We will classify the eigenvalues of the Pomeranchuk channel in terms of the irreducible representations of the point group $D_3$.
   There are two one-dimensional representations $A_1$ and $A_2$, and one two-dimensional representation $E$~\cite{Hamermesh}.
   Furthermore, we find two distinct eigenvalues (each doubly degenerate) that belong to the representation $E$, and we label them by $E^+$ and $E^{-}$.
   To connect to the commonly used notation of continuous rotation symmetry, note that one could assign
$s$-wave symmetry to the $A_1$ representation, $d$-wave and $g$-wave  to the $E$ representation, for our choice of $E^-$ and $E^+$, and $f$-wave to the $A_2$ representation.
    The irreducible representations also contain harmonics of higher order.
   The eigenvalues for the spin and charge  Pomeranchuk channel are identical for $E$ and $A_2$ representations:
\begin{equation}
\begin{gathered}
\lambda_{C Pom}^{E^-} = \lambda_{SPom}^{E^-} = \Pi_{++}(0) \left(u -g_2 - \sqrt{g_{1+}^2 - g_{1+} g_{1-} + g_{1-}^2 - g_{1-} g_3 - g_{1+} g_3 + g_3^2 }  \right), \\
\lambda_{C Pom}^{E^+} = \lambda_{SPom}^{E^+} = \Pi_{++}(0) \left(u -g_2 + \sqrt{g_{1+}^2 - g_{1+} g_{1-} + g_{1-}^2 - g_{1-} g_3 - g_{1+} g_3 + g_3^2 }  \right), \\
\lambda_{C Pom}^{A_2} =  \lambda_{SPom}^{A_2} = \Pi_{++}(0) \left( -g_{1+} - g_{1-} + 2 g_2 - g_3 + u  \right),
\end{gathered}
\label{12Pom_c_eig}
\end{equation}
but differ for the $A_1$ representation
\begin{equation}
\begin{gathered}
\lambda_{C Pom}^{A_1} = \Pi_{++}(0) \left( g_{1+} + g_{1-} + 2 g_2 + g_3 - 11 u  \right), \\
\lambda_{SPom}^{A_1} = \Pi_{++}(0) \left( g_{1+} + g_{1-} + 2 g_2 + g_3 + u  \right).
\end{gathered}
\label{12Pom_a1}
\end{equation}
Using Eq. (\ref{coupl_2}) for the dependence of the couplings on the parameter $\alpha_T$, we obtain
the eigenvalues as  functions of $\alpha_T$. We plot them in Fig. \ref{lambda_alpha_12p} a.
 The coupling in the charge $A_1$ channel is strongly repulsive, but the one in the spin $A_1$ channel is attractive.
 For $A_2$ and $E$ representations, the eigenvalues are attractive, and the strongest one is in the $E^+$ channel ($d-$wave Pomeranchuk), see Fig. \ref{lambda_alpha_12p} a.

Comparing the magnitudes of the eigenvalues in different Pomeranchuk channels, we find that the strongest attraction is in the $A_1$ spin Pomeranchuk channel. The attraction in this channel holds when $\alpha_T =0$, and increases with $\alpha_T$.  Note that  the subleading $E_+$
($g$-wave)
spin/charge channel is also attractive at $\alpha_T =0$, and  the attraction increases with $\alpha_T$.
Its counterpart $E^-$ ($d$-wave) is also attractive, but with decreasing attraction for increasing $\alpha_T$.
This
situation is more complex than in
 the six-patch model, where the  attraction in the $d$-wave Pomeranchuk channel decreases with $\alpha_T$.

We next analyze the eigenvalues in the  density wave channels. For the intra-valley density-wave channels, we can read off the eigenvalues from Eqs.~\eqref{eq:CDWQ3}-\eqref{eq:CDWQ11}
for CDW
\begin{equation}
\begin{gathered}
\lambda_{CDW} (\vec{Q}_3) =  \Pi_{++}\left(\vec{Q}_{3}\right) \left( u -2 g_{1-} \right), \\
\lambda_{CDW} (\vec{Q}_4) =  \Pi_{++}\left(\vec{Q}_{4}\right) \left( u -2 g_{2} \right), \\
\lambda_{CDW} (\vec{Q}_7) =  \Pi_{++}\left(\vec{Q}_{7}\right) \left( u -2 g_{3} \right), \\
\lambda_{CDW} (\vec{Q}_{11}) =  \Pi_{++}\left(\vec{Q}_{11}\right) \left( u -2 g_{1+} \right),
\end{gathered}
\end{equation}
and
from Eq.~\ref{eq:intraSDW}
for the intra-valley SDW channels
\begin{equation}
\lambda_{SDW} =  \Pi_{++}\left(\vec{Q}_{3,4,7,11}\right) u,
\end{equation}
We use Eq. (\ref{coupl_2}) for the couplings and Table \ref{polar_table_12p} for the polarization bubbles and obtain $\lambda_{CDW}$ and $\lambda_{SDW}$ at various $\vec{Q}$ as functions of $\alpha_T$. We plot the results in
 Fig. \ref{lambda_alpha_12p} b for CDW and in Fig. \ref{lambda_alpha_12p} c for SDW channels.

For inter-valley channels, the eigenvalues in SDW and CDW sub-channels are still degenerate for a given momentum transfer, but there are two possible eigenvalues for every block in the block-diagonal matrix. We label the eigenvalues  with the superscript $+/-$, corresponding to  $(1,\pm 1)$ within every block.
The eigenvalues are given by
\begin{equation}
\begin{gathered}
\lambda_{CDW}^{+} (\vec{Q}_1) = \lambda_{SDW}^{+} (\vec{Q}_1) = \Pi_{+-}\left(\vec{Q}_{1}\right) \left( u + g_3 \right); \\
\lambda_{CDW}^{-} (\vec{Q}_1) = \lambda_{SDW}^{-} (\vec{Q}_1)  =  \Pi_{+-}\left(\vec{Q}_{1}\right) \left( u - g_3 \right);
\\
\lambda_{CDW}^{+} (\vec{Q}_2) = \lambda_{SDW}^{+} (\vec{Q}_2)  = \Pi_{+-}\left(\vec{Q}_{2}\right) \left( u + g_2 \right); \\
\lambda_{CDW}^{-} (\vec{Q}_2) = \lambda_{SDW}^{-} (\vec{Q}_2)  =  \Pi_{+-}\left(\vec{Q}_{2}\right) \left( u - g_2 \right);
\\
\lambda_{CDW}^{+} (\vec{Q}_5) = \lambda_{SDW}^{+} (\vec{Q}_5)  = \Pi_{+-}\left(\vec{Q}_{5}\right) \left( u + g_{1+} \right); \\
\lambda_{CDW}^{-} (\vec{Q}_5) = \lambda_{SDW}^{-} (\vec{Q}_5)  =  \Pi_{+-}\left(\vec{Q}_{5}\right) \left( u - g_{1+} \right);
\\
\lambda_{CDW}^{+} (\vec{Q}_9) = \lambda_{SDW}^{+} (\vec{Q}_9) = \Pi_{+-}\left(\vec{Q}_{9}\right) \left( u + g_{1-} \right); \\
\lambda_{CDW}^{-} (\vec{Q}_9) = \lambda_{SDW}^{-} (\vec{Q}_9) =  \Pi_{+-}\left(\vec{Q}_{9}\right) \left( u - g_{1-} \right);
\end{gathered}
\end{equation}
For the inter-valley channel with momentum transfer $\vec{Q}_{6}$, there is one eigenvalue per block.
  The eigenvalue for this channel is
\begin{equation}
\lambda_{CDW} (\vec{Q}_6) = \lambda_{SDW} (\vec{Q}_6) = \Pi_{+-}\left(\vec{Q}_{6}\right) u
\end{equation}
We plot the eigenvalues as functions of $\alpha_T$ in Fig. \ref{lambda_alpha_12p} d for the $+$ channels  and in Fig. \ref{lambda_alpha_12p} e for the $-$ channels. The eigenvalue for the channel with $\vec{Q}_6$ is shown in Fig. \ref{lambda_alpha_12p} f.

\begin{figure}[t]
\center{\includegraphics[width=0.4\linewidth]{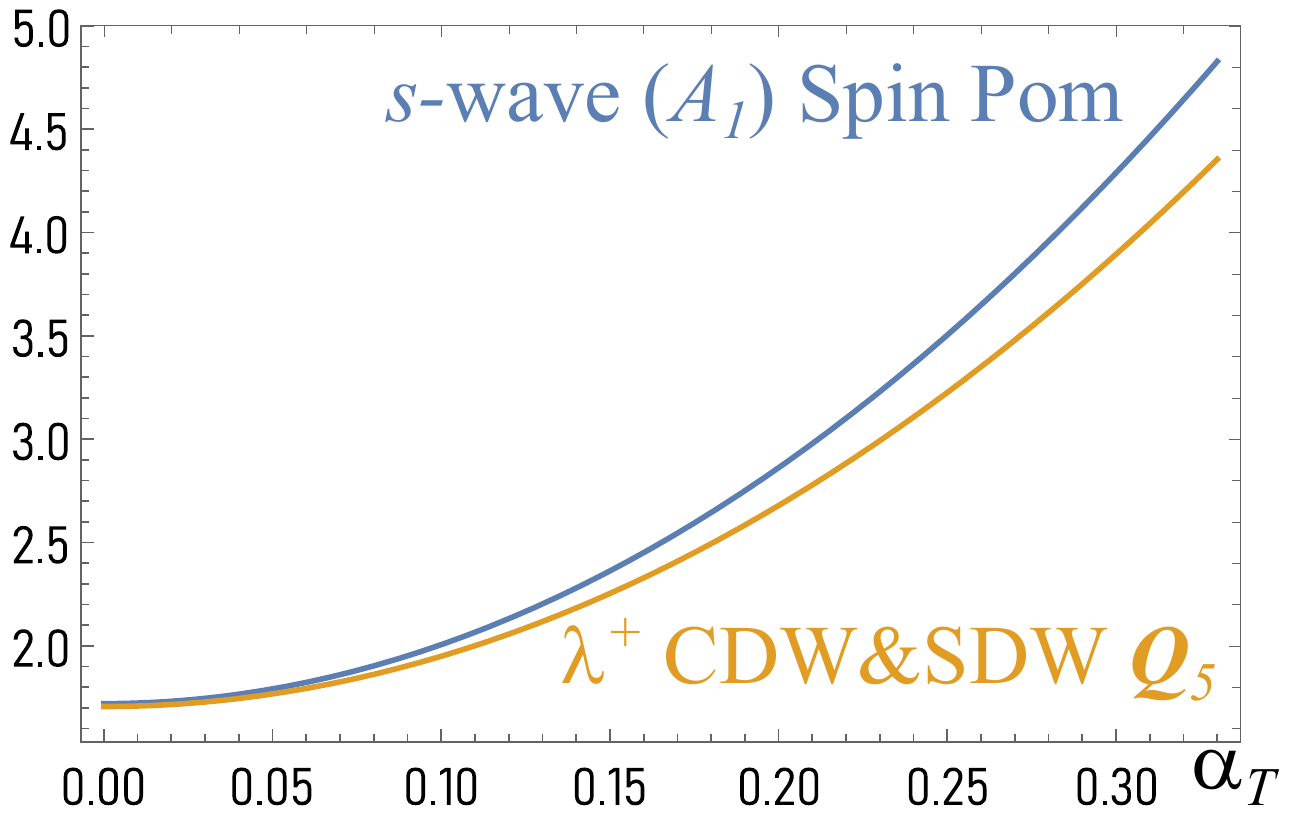} }
\centering{}\caption{The two largest positive eigenvalues for the 12-patch model, as  functions of $\alpha_T$.
The largest eigenvalue for all $\alpha_T$ is in the $s-$wave spin Pomeranchuk channel.  An instability in this channel  gives rise to FM/AFM order (intra-valley FM/inter-valley AFM).}
\label{12p_lead}
\end{figure}

We now compare the eigenvalues in the Pomeranchuk channels and intra-valley and inter-valley CDW/SDW channels.
In Fig. \ref{12p_lead}   we show the two most strongly attractive couplings as functions of $\alpha_T$.
  We find that for the twelve-patch model the largest coupling is in the $s-$wave spin Pomeranchuk channel .
    The corresponding eigenvalue $\lambda_{Pom,s}^{A_1}$ is double degenerate, reflecting that at this level of consideration, an $s-$wave spin Pomeranchuk order introduces two ferromagnetic orders, one per valley.
     This is quite similar to what we found earlier for the six-patch model.
     Like there, the relative orientation of the two ferromagnetic orders is set by the coupling to fluctuating CDW/SDW order parameters  with momentum $\vec{Q}_5$, for which the eigenvalue is
      second largest. These CDW/SDW order parameters involve fermions from different valleys and provide an effective interaction  between ferromagnetic order parameters on different valleys. The  free energy functional has the same form as in the six-patch case, and like there,
       CDW/SDW fluctuations select antiparallel orientation of ferromagnetic orders on the two valleys.
        As a result, the order parameter is again FM/AFM -- ferromagnetic within a valley and antiferromagnetic between the valleys.
        The only difference with the six patch model is that now the coupling in this channel is the strongest one for all $\alpha_T$.

The free energy for the
$g$-wave ($E^+$) charge and spin Pomeranchuk order parameters also has the same form as in the six-patch model
 because they both belong to the same irreducible representation $E$.
This means that the ordered state is a nematic -- it breaks lattice rotational symmetry. Even if this order does not develop, the attraction gives rise to enhanced nematic fluctuations.
We note in passing that in the particle-particle channel, the attractive interaction is in the
$E$ ($g-$wave) and $A_2$ ($i-$wave)
 channels~\cite{Chichinadze2019nematic}.

\section{Conclusions}
\label{sec:Concl}

In this work we continued our analysis of the  effects of interactions  in twisted bilayer graphene
near Van Hove filling taking into account the special non-local form of the interactions.
We emphasize in this regard
  that two recent theoretical studies\cite{PhysRevB.100.205113,PhysRevB.100.205114}
  found that long-range electrostatic
   interactions pin Van Hove singularities to the Fermi level for a broad range of fillings, and
   that
   recent experimental evidence for
   indicates the presence of
    multiple Van Hove singularities for doping values near $n = \pm 2$ and $n = \pm 3$. \cite{wu2020chern}.
In our previous work~\cite{Chichinadze2019nematic} we studied the interactions in the particle-particle channel, which give rise to superconductivity, and argued that a superconducting order can also break lattice rotational symmetry
(a nematic superconductor). In this paper, we reported the results of our analysis of the effects of interactions in the particle-hole channel. An instability in a particle-hole channel can give rise to SDW, CDW, ferro/antiferromagnetism, and a nematic order, which compete with superconductivity.
 We identified particle-hole channels with the largest attractive interactions and analyzed the structure of the corresponding order parameters.

The point of departure for our analysis is an effective patch model for itinerant interacting fermions near Van Hove points.
 The density of states near Van Hove points is singular, and this enhances the strength of the
 interaction effects.
 We argued that twisted bilayer graphene can have either six or twelve Van Hove points, depending on the details of the electronic dispersion, and studied both six-patch and twelve-patch models.
    We included all possible interactions between low-energy fermions in the patches and used the real-space microscopic interaction Hamiltonian, suggested by Kang and Vafek~\cite{Kang2018strong}, to obtain the relative magnitudes of these interactions. The Hamiltonian consists of  a cluster Hubbard term, which contains density-density interactions between sites of a given hexagon in the moiré lattice, and a term with bilinear combinations of hoppings between different sites of a hexagon.
 The relative strength of the second term is  specified by the parameter  $\alpha_T$, and we analyze the interplay between couplings in different particle-hole channels as a function of $\alpha_T$.

 There are three main results reported in this paper.
  First, we find
   the intra-valley  ferromagnetism
   as the leading instability  for any $\alpha_T$ in the twelve-patch model and for large enough $\alpha_T$ in the six-patch model.  In both models,  the magnitudes of the ferromagnetic order parameters in the two valleys are equal, but their relative orientation is determined by subleading effects.  We found that
   inter-valley hopping
   terms favor antiferromagnetic ordering between the valleys, i.e. FM/AFM order  (intra-valley FM/inter-valley AFM).
     The same  has been found in the strong coupling limit ~\cite{Kang2018strong,PhysRevB.100.205131,PhysRevResearch.2.013370}.
     We caution, however, that in TBG inter-valley mixing terms are believed to be small and may potentially be
     smaller than
     subleading terms, like the
     cubic coupling
     of
     ferromagnetic
      and
      degenerate CDW and SDW  fluctuations, mediated by fermions outside of Van Hove regions. This last coupling favors a FM ordering between the valleys.

Second, we find a highly non-trivial mixed CDW/SDW order with $120^\circ$ orientation of three vector SDW components of the order parameter.
The order parameters are complex due to the valley degree of freedom of twisted bilayer graphene, and we find that the relative phase between CDW and SDW order is $\pm\pi/2$.
This order competes with FM/AFM, particularly in the six-patch model.

  Third, in both six-patch and twelve-patch models, we found an attraction in $d-$wave spin and charge Pomeranchuk channels.  The attraction holds even when $\alpha_T = 0$,  due  to the
non-local
nature of the density-density interaction. We argued that
charge and spin Pomeranchuk orders are degenerate in the absence of valley mixing, and that
an instability in
one of these channels (or both)
breaks the lattice rotational symmetry,
i.e.
 gives rise to a nematic order.  In our calculations, the couplings in the nematic channels are subleading to those in FM/AFM and CDW/SDW channels.  Still, the very fact that the nematic couplings are attractive implies that there should be sizable nematic fluctuations. This agrees with the results of STM studies~\cite{stm2019,Cao2020nematicity}.
Overall, our results show that the physics near Van Hove filling is quite rich and includes not only
 superconductivity, but also competing orders/strong fluctuations in the particle-hole channel.

The competition between superconductivity and different particle-hole instabilities depends on microscopic details. We find parameter regions (for small $\alpha_T$), where the pairing interaction is repulsive, but the interaction in the particle-hole channel is attractive. When both particle-particle and particle-hole channels are attractive, the superconducting instability wins in the limit of weak coupling, because the particle-particle bubble scales like $\log^2T$, while the particle-hole bubble only scales like $\log T$. At stronger couplings, however, the instabilities develop at higher temperatures, and the charge or spin orders that we found can overcome superconductivity.

\section{Acknowledgments}
We thank V. Andreeva, E. Andrei, M. Christensen, R. Fernandes, L. Fu, D. Goldhaber-Gordon, P. Jarillo-Herrero, J. Kang,  A. Klein, J. Schmalian, D. Shaffer, O. Vafek, and A. Vishwanath for fruitful discussions.
 The work was supported by  U.S. Department of Energy, Office of Science, Basic Energy Sciences, under Award No. DE-SC0014402.
D.C. gratefully acknowledges support from the Allen M. Goldman Fellowship of the University of Minnesota. L.C. was supported by the Humboldt foundation, and work at BNL is supported by the U.S. Department of Energy (DOE), Office of Basic Energy Sciences, under Contract No. DE-SC0012704.

\begin{appendix}

\section{Technical details of Hubbard-Stratonovich transformation}
\label{app:HStechs}

To perform the Hubbard-Stratonovich transformation
in the six-patch model
we introduce matrices of Green's function and test fields
for SDW ($\vec M_i$), CDW ($\Delta_i$), spin Pomeranchuk ($\vec S_i$) and charge Pomeranchuk ($C_i$)

\begin{equation}
\begin{gathered}
\hat{G}_0 = \begin{pmatrix}
G_1\mathbbm{1} & 0 & 0 & 0 & 0 & 0 \\
0 & G_2 \mathbbm{1}  & 0 & 0 & 0 & 0 \\
0 & 0 & G_3 \mathbbm{1}  & 0 & 0 & 0 \\
0 & 0 & 0 & G_{1'} \mathbbm{1}  & 0 & 0 \\
0 & 0 & 0 & 0 & G_{2'} \mathbbm{1}  & 0 \\
0 & 0 & 0 & 0 & 0 & G_{3'} \mathbbm{1}
\end{pmatrix}, \;
\hat{\Delta} = \begin{pmatrix}
0 & 0 & 0 & 0 & \bar{\Delta}_3 \mathbbm{1} & \bar{\Delta}_2  \mathbbm{1} \\
0 & 0 & 0 & \bar{\Delta}_3  \mathbbm{1} & 0 & \bar{\Delta}_1 \mathbbm{1} \\
0 & 0 & 0 & \bar{\Delta}_2 \mathbbm{1} & \bar{\Delta}_1 \mathbbm{1} & 0 \\
0 & \Delta_3 \mathbbm{1} & \Delta_2  \mathbbm{1} & 0 & 0 & 0 \\
\Delta_3 \mathbbm{1} & 0 & \Delta_1 \mathbbm{1} & 0 & 0 & 0 \\
\Delta_2 \mathbbm{1} & \Delta_1 \mathbbm{1} & 0 & 0 & 0 & 0
\end{pmatrix}, \\
\hat{M} = \begin{pmatrix}
0 & 0 & 0 & 0 & \bar{\vec{M}}_3 \cdot \vec{\sigma} & \bar{\vec{M}}_2 \cdot \vec{\sigma} \\
0 & 0 & 0 & \bar{\vec{M}}_3 \cdot \vec{\sigma} & 0 & \bar{\vec{M}}_1 \cdot \vec{\sigma} \\
0 & 0 & 0 & \bar{\vec{M}}_2 \cdot \vec{\sigma} & \bar{\vec{M}}_1 \cdot \vec{\sigma} & 0 \\
0 & \vec{M}_3 \cdot \vec{\sigma} & \vec{M}_2 \cdot \vec{\sigma} & 0 & 0 & 0 \\
\vec{M}_3 \cdot \vec{\sigma} & 0 & \vec{M}_1 \cdot \vec{\sigma} & 0 & 0 & 0 \\
\vec{M}_2 \cdot \vec{\sigma} & \vec{M}_1 \cdot \vec{\sigma} & 0 & 0 & 0 & 0
\end{pmatrix}, \;
\hat{S} = \begin{pmatrix}
\vec{S}_1 \cdot \vec{\sigma} & 0 & 0 & 0 & 0 & 0 \\
0 & \vec{S}_2 \cdot \vec{\sigma}  & 0 & 0 & 0 & 0 \\
0 & 0 & \vec{S}_3 \cdot \vec{\sigma}  & 0 & 0 & 0 \\
0 & 0 & 0 & \vec{S}_{1'} \cdot \vec{\sigma}  & 0 & 0 \\
0 & 0 & 0 & 0 & \vec{S}_{2'} \cdot \vec{\sigma}  & 0 \\
0 & 0 & 0 & 0 & 0 & \vec{S}_{3'} \cdot \vec{\sigma}
\end{pmatrix},\\
\hat{C} = \begin{pmatrix}
C_1\mathbbm{1}    & 0 & 0 & 0 & 0 & 0 \\
0 &C_2 \mathbbm{1}  & 0 & 0 & 0 & 0 \\
0 & 0 & C_3\mathbbm{1}    & 0 & 0 & 0 \\
0 & 0 & 0 & C_4  \mathbbm{1}  & 0 & 0 \\
0 & 0 & 0 & 0 & C_5  \mathbbm{1}  & 0 \\
0 & 0 & 0 & 0 & 0 & C_6\mathbbm{1}
\end{pmatrix},
\end{gathered}
\end{equation}
where
\begin{equation}
\vec{B} \cdot \vec{\sigma} = \begin{pmatrix}
B_z & B_x - i B_y \\
B_x + i B_y & -B_z
\end{pmatrix},
\end{equation}
with
$\vec{B} = (B_x,B_y,B_z),$ and the  vector of Pauli matrices $\vec{\sigma}$.
After performing the Hubbard-Stratonovich transformation,
the fermionic Hamiltonian is of the form
\beq
H=\psi^\dagger (G_0^{-1}+\hat\Delta+\hat M+\hat S+\hat C)\psi\,,
\eeq
where $\psi^\dagger=(f^\dagger_1,f^\dagger_2,f^\dagger_3,f^\dagger_{1'},f^\dagger_{2'},f^\dagger_{3'})$ and $f^\dagger_i=(f^\dagger_{i\uparrow},f^\dagger_{i,\downarrow})$.
Integrating out fermions, we get
\begin{equation}
\mathrm{Tr} \; \mathrm{ln} \; \hat{G}^{-1} = \mathrm{Tr} \; \mathrm{ln} \left( \hat{G}_0^{-1} (1 + \hat{G}_0 (\hat{\Delta} + \hat{M} + \hat{S} +\hat C))  \right) =
\mathrm{const} + \mathrm{Tr} \; \mathrm{ln} \left( 1 + \hat{G}_0 (\hat{\Delta} + \hat{M} + \hat{S} +\hat C)  \right),
\end{equation}
where the trace is taken over patch space and spin space.
Before taking the trace, we replace $\vec S_i=\vec S$ for $s$-wave order or by $\vec S_1=\frac{2}{\sqrt{6}}\vec\phi_{d2}$, $\vec S_{2/3}=\frac{1}{\sqrt{2}}\vec\phi_{d1}\mp\frac{1}{\sqrt{6}}\vec \phi_{d2}$ and $C_1=\frac{2}{\sqrt{6}}\chi_{d2}$, $C_{2/3}=\frac{1}{\sqrt{2}}\chi_{d1}\mp\frac{1}{\sqrt{6}}\chi_{d2}$ for $d$-wave order (analogously for $\vec S_{i'}$ and $C_{i'}$).
We now expand the $\mathrm{log}$ in small Hubbard-Stratonovich fields $\hat{\Delta}, \hat{M},$ $\hat{S}$ and $\hat C$. In the quadratic order one only gets non-mixed terms, i.e.
\begin{equation}
\begin{gathered}
\mathrm{Tr} [\hat{G}_0 \hat{\Delta} \hat{G}_0 \hat{\Delta} ] = \Pi(\vec{Q}_s) \sum_i |\Delta_i|^2, \\
\mathrm{Tr} [\hat{G}_0 \hat{M} \hat{G}_0 \hat{M} ] = \Pi(\vec{Q}_s) \sum_i (\bar{\vec{M}}_i \cdot \vec{M}_i), \\
\end{gathered}
\end{equation}
and
\beq
\mathrm{Tr} [\hat{G}_0 \hat{S} \hat{G}_0 \hat{S} ]
\text{ or } \mathrm{Tr} [\hat{G}_0 \hat{C} \hat{G}_0 \hat{C} ],
\eeq
because of momentum conservation and since the Pauli matrices obey $\mathrm{Tr} [ \sigma ] = 0, \mathrm{Tr} [ \sigma_i \sigma_j ] = 2 \delta_{ij},$ where $\sigma_i$ are Pauli matrices, $\delta_{ij}$ is the Kronecker symbol, and $\Pi(\vec{k})$ is the polarization operator with transferred momentum $\vec{k}$. Typically, odd-order terms (like cubic) vanish upon taking the trace.
However,
in our case this type of terms can be allowed
by symmetry.
Expanding the log to third order we get two different cubic terms
contributing to the free energy for CDW/SDW and FM/AFM fields
\begin{equation}
\begin{gathered}
\mathrm{Tr} [\hat{G}_0 \hat{\Delta} \hat{G}_0 \hat{M} \hat{G}_0 \hat{S} ]
\end{gathered}
\end{equation}
which leads to the first term in Eq.~\ref{eq:coupl3}.
The cubic terms in the free energy for $d$-wave charge and spin Pomeranchuk fluctuations is obtained from
\begin{align}
\mathrm{Tr} [\hat{G}_0 (\hat{C}+\hat S) \hat{G}_0 (\hat{C}+\hat S) \hat{G}_0 (\hat{C}+\hat S) ]
\end{align}
with the result given in Eq.~\ref{eq:FCSPom}.

 We now proceed to the quartic terms
 i.e. we expand the logarithm to quartic order in the fields.
 In the case of CDW/SDW and FM/AFM order, we can use the following simplifications for products that contain only two different fields:
 the trace over an odd number of  $\hat{\Delta}$ matrices gives zero and due to the
 resulting odd number of Pauli matrices.
 Traces with odd number of $\hat{S}$ or $\hat{M}$ vanish due to the momentum conservation constraints (there are no such possible square box diagrams). For example,
\begin{equation}
\mathrm{Tr} [\hat{G}_0 \hat{S} \hat{G}_0 \hat{S} \hat{G}_0 \hat{S} \hat{G}_0 \hat{M} ] =  \mathrm{Tr} [\hat{G}_0 \hat{\Delta} \hat{G}_0 \hat{\Delta} \hat{G}_0 \hat{S} \hat{G}_0 \hat{M} ] = 0.
\end{equation}
There is one combination
 (plus its cyclic permutations), which couples all three fields
\begin{equation}
\begin{gathered}
\mathrm{Tr} [\hat{G}_0 \hat{\Delta} \hat{G}_0 \hat{S} \hat{G}_0 \hat{M} \hat{G}_0 \hat{S} ] = - 2 i Z_1 \biggr[ \left( \bar{\Delta}_1 \vec{M}_1 -\Delta_1  \bar{\vec{M}}_1 \right) \left(\vec{S}_{3'}  \times  \vec{S}_{2} + \vec{S}_{2'} \times \vec{S}_{3}  \right)  \\
+ \left( \bar{\Delta}_2 \vec{M}_2 -\Delta_2  \bar{\vec{M}}_2 \right) \left( \vec{S}_{3'} \times \vec{S}_{1} + \vec{S}_{1'} \times \vec{S}_{3}  \right) +
\left( \bar{\Delta}_3 \vec{M}_3 -\Delta_3  \bar{\vec{M}}_3 \right) \left( \vec{S}_{1'} \times \vec{S}_{2} + \vec{S}_{2'} \times \vec{S}_{1}  \right)  \biggr].
\end{gathered}
\end{equation}
Here, we used $\mathrm{Tr}  [ \sigma_i \sigma_j \sigma_k] = 2 i \epsilon_{ijk}$, where $\epsilon_{ijk}$ is the Levi-Civita tensor.
For the other terms using the invariance of the trace operation under the cyclic permutation of matrices in the product we expand and get
\begin{equation}
\begin{gathered}
F_4 = F_4^{\Delta} + F_4^{M} + F_4^{S} + F_4^{\Delta,M} + F_4^{\Delta,S} + F_4^{M,S}, \\
F_4^{\Delta} = \mathrm{Tr} [\hat{G}_0 \hat{\Delta} \hat{G}_0 \hat{\Delta} \hat{G}_0 \hat{\Delta} \hat{G}_0 \hat{\Delta} ], \\
F_4^{M} = \mathrm{Tr} [\hat{G}_0 \hat{M} \hat{G}_0 \hat{M} \hat{G}_0 \hat{M} \hat{G}_0 \hat{M} ], \\
F_4^{S} = \mathrm{Tr} [\hat{G}_0 \hat{S} \hat{G}_0 \hat{S} \hat{G}_0 \hat{S} \hat{G}_0 \hat{S} ], \\
F_4^{\Delta,M} = 4 \mathrm{Tr} [\hat{G}_0 \hat{\Delta} \hat{G}_0 \hat{\Delta} \hat{G}_0 \hat{M} \hat{G}_0 \hat{M} ] + 2 \mathrm{Tr} [\hat{G}_0 \hat{\Delta} \hat{G}_0 \hat{M} \hat{G}_0 \hat{\Delta} \hat{G}_0 \hat{M} ], \\
F_4^{\Delta,S} = 4 \mathrm{Tr} [\hat{G}_0 \hat{\Delta} \hat{G}_0 \hat{\Delta} \hat{G}_0 \hat{S} \hat{G}_0 \hat{S} ] + 2 \mathrm{Tr} [\hat{G}_0 \hat{\Delta} \hat{G}_0 \hat{S} \hat{G}_0 \hat{\Delta} \hat{G}_0 \hat{S} ], \\
F_4^{M,S} = 4 \mathrm{Tr} [\hat{G}_0 \hat{M} \hat{G}_0 \hat{M} \hat{G}_0 \hat{S} \hat{G}_0 \hat{S} ] + 2 \mathrm{Tr} [\hat{G}_0 \hat{M} \hat{G}_0 \hat{S} \hat{G}_0 \hat{M} \hat{G}_0 \hat{S} ].
\end{gathered}
\end{equation}
Further evaluating traces we obtain the free energy shown in the main text.

Terms quartic in the  $d$-wave charge and spin Pomeranchuk fields are obtained from
\begin{align}
&\mathrm{Tr} [\hat{G}_0 \hat S \hat{G}_0 \hat S \hat{G}_0 \hat S \hat{G}_0 \hat S]\\
&\mathrm{Tr} [\hat{G}_0 \hat{C} \hat{G}_0 \hat{C}\hat{G}_0 \hat{C}\hat{G}_0 \hat{C}]\\
&\mathrm{Tr} [\hat{G}_0 \hat{C} \hat{G}_0 \hat S \hat{G}_0 \hat{C} \hat{G}_0 \hat S] =\mathrm{Tr} [\hat{G}_0 \hat{S} \hat{G}_0 \hat S \hat{G}_0 \hat{C} \hat{G}_0 \hat C] \\
\end{align}
and permutations thereof. Again products odd in the fields vanish because $\mathrm{Tr}\sigma=0$ or because the product $\vec\phi_i\cdot(\vec\phi_j\times \vec\phi_k)=0$ with the only two possibilities for the vectors $\vec\phi_{d1}$ and $\vec\phi_{d2}$.
\end{appendix}

\bibliography{biblio}

\end{document}